\documentclass[longauth]{aa}  

\usepackage{graphicx}
\usepackage{tablefootnote}
\usepackage{threeparttable}
\usepackage{float}
\usepackage{setspace}
\usepackage{subcaption}

\usepackage[dvipsnames]{xcolor}
\usepackage[colorlinks=true,citecolor=blue,urlcolor=blue,anchorcolor = blue,linkcolor = WildStrawberry]{hyperref}

\usepackage{txfonts}

\usepackage{soul}
\usepackage{xcolor}
\usepackage{amsmath}

\usepackage{supertabular,booktabs}

\usepackage{caption}
\usepackage{subcaption}
\usepackage{longtable}
\usepackage{rotating}
\usepackage{lscape}
\usepackage{amsmath}

\newcommand{\msun}{$M_{\odot}$}
\newcommand{\rsun}{$R_{\odot}$}
\newcommand{\lsun}{$L_{\odot}$}
\newcommand{\vrad}{$v_\mathrm{rad}$}
\newcommand{\teff}{$T_\mathrm{eff}$}
\newcommand{\logg}{$\log g$}
\newcommand{\logy}{$\log n(\mathrm{He}) / n(\mathrm{H})$}
\newcommand{\vsini}{$v_\mathrm{rot}\sin i$}

\newcommand{\atlas}{\textsc{Atlas}{\footnotesize12}}
\newcommand{\detail}{\textsc{Detail}}
\newcommand{\surface}{\textsc{Surface}}
\newcommand{\kms}{km\,s$^{-1}$}

\begin{document} 

   \title{A 500 pc volume-limited sample of hot subluminous stars}
   \subtitle{II. Atmospheric parameters, mass distribution, and kinematics}
   \author{H. Dawson
          \inst{1}
          \and
          M. Dorsch\inst{1}
          \and
          S. Geier\inst{1}
          \and
          J. Munday \inst{2}
          \and
          M. Pritzkuleit\inst{1}
          \and
          U. Heber\inst{3}
          \and
          F. Mattig \inst{1}
          \and          
          D. Benitez-Palacios\inst{4}
          \and
          M. {Vu{\v{c}}kovi{\'c}}\inst{4} 
          \and
          I. Pelisoli\inst{2,1}
          \and
          K. Deshmukh\inst{6,19}
          \and
          A. Bhat \inst{1}
          \and
          L. Kufleitner \inst{3}
          \and
          M. Uzundag\inst{4,5,6}
          \and
          V.~Schaffenroth \inst{7,1}
          \and
          N. Reindl \inst{8, 1}
          \and
          R. Culpan\inst{1}  
          \and
          R. Raddi \inst{9}
          \and
          L. Antunes Amaral \inst{4,5}
          \and
          A. G. Istrate \inst{10}
          \and
          S. Justham \inst{11} 
          \and
          R. H. Østensen \inst{12}
          \and
          J. H. Telting \inst{13,14}
          \and
          T. Steinmetz \inst{15,16,17}
          \and
          N. Rodríguez-Segovia \inst{18}
          \and
          P. Fernandez-Schlosser \inst{4}
          \and
          A. Durán-Reyes \inst{4}
          \and
          E. Arancibia-Rojas \inst{4}
          \and
          M. Latour \inst{20}
          \and
          G. T. Jones \inst{2}
          \and
          M. O'Brien\inst{2}
          \and
          S. Sahu\inst{2}
          \and
          A. Elms\inst{2}
}   
          
   \institute{Institute for Physics and Astronomy, University of Potsdam, Karl-Liebknecht-Str. 24/25, 14476 Potsdam, Germany\\
              \email{hdawson@astro.physik.uni-potsdam.de}
         \and
            Department of Physics, University of Warwick, Gibet Hill Road, Coventry CV4 7AL, UK
         \and
            Dr. Remeis-Sternwarte and ECAP, Astronomical Institute, University of Erlangen-Nürnberg, Sternwartstr. 7, D-96049 Bamberg, Germany
         \and
             Instituto de Física y Astronomía, Universidad de Valparaíso, Gran Bretaña 1111, Playa Ancha, Valparaíso 2360102, Chile
         \and
             European Southern Observatory, Alonso de Cordova 3107, Santiago, Chile
         \and
            Institute of Astronomy, KU Leuven, Celestijnenlaan 200D, B-3001 Leuven, Belgium 
         \and
            Thüringer Landessternwarte Tautenburg, Sternwarte 5, D-07778 Tautenburg, Germany 
         \and
            Landessternwarte Heidelberg, Zentrum fur Astronomie, Ruprecht-Karls-Universitat, Konigstuhl 12, 69117 Heidelberg, Germany
         \and
             Universitat Politècnica de Catalunya, Departament de Física, c/ Esteve Terrades 5, 08860 Castelldefels, Spain 
         \and
             Department of Astrophysics/IMAPP, Radboud University, P O Box 9010, NL-6500 GL Nijmegen, The Netherlands 
         \and
             Max Planck Institut für Astrophysik, Karl-Schwarzschild-Straße 1, 85748 Garching bei München, Germany
        \and
            Recogito AS, Storgaten 72, N-8200 Fauske, Norway 
         \and
             Nordic Optical Telescope, Rambla José Ana Fernández Pérez 7, ES-38711 Breña Baja, Spain 
        \and
            Department of Physics and Astronomy, Aarhus University, Munkegade 120, DK-8000 Aarhus C, Denmark 
        \and
            Isaac Newton Group of Telescopes, Apartado de Correos 368, E-38700 Santa Cruz de La Palma, Spain
        \and
            Department of Physics and Astronomy, University of Sheffield, Sheffield S3 7RH, UK 
        \and
            Nicolaus Copernicus Astronomical Centre, ul. Rabia\'{n}ska 8, 87-100 Toru\'{n}, Poland 
        \and
            School of Science, University of New South Wales, Australian Defence Force Academy, Canberra, ACT 2600, Australia 
        \and
            Leuven Gravity Institute, KU Leuven, Celestijnenlaan 200D, box 2415, 3001 Leuven, Belgium 
        \and
            Institut für Astrophysik und Geophysik, Georg-August-Universität Göttingen, Friedrich-Hund-Platz 1, 37077 Göttingen, Germany
}

   \date{Received November 14, 2025 / Accepted January 4, 2026}

  \abstract
{
We present a quantitative spectroscopic and kinematic analysis of a volume-complete sample of hot subluminous stars within 500~pc of the Sun, assembled using accurate parallax measurements from the \textit{Gaia} space mission data release 3 (DR3).
In total, 3226 spectra of 253 hot subdwarf stars were analysed to derive atmospheric parameters ($T_\mathrm{eff}$, $\log g$, and helium abundance) and radial velocities. 
Spectral energy distributions (SEDs) with \textit{Gaia} parallaxes were used to measure stellar radii, luminosities, and masses.
The derived atmospheric parameters reveal a consistent alignment between sdB and sdO stars in the Kiel diagram when compared to theoretical evolutionary models.
Notably, we find a substantial population (about 10\%) of hot subdwarfs located below the $0.45$\,\msun\ zero-age EHB in both the Kiel and Hertzsprung-Russell diagrams (HRD), which likely originate from intermediate-mass progenitors (1.8 to 8\,$M_\odot$). 
The overall mass distribution peaks at $0.48^ {+0.14}_{-0.10}$$M_\odot$, while hot subdwarfs below the EHB peak at $0.43^ {+0.11}_{-0.09}$ $M_{\odot}$, supporting a scenario of non- or semi-degenerate helium ignition for these objects characteristic of intermediate-mass stars. Interpolation of EHB and post-EHB tracks yields theoretical mass distributions consistent with our estimates based on SED and parallax. By assuming a mass range between 0.40 and 0.50 $M_{\odot}$ during the interpolation, we further find that the post-EHB birthrate in our sample is 2-3 times higher than the EHB birthrate, which may suggest overestimated EHB lifetimes in theoretical tracks or contamination from other formation and evolutionary channels. 
Our kinematic analysis shows that $86\pm2$\% of the stars belong to the Galactic thin disk, with $13\pm1$ \% and $1\pm1$\% associated with the thick disk and halo, respectively. The below-EHB population is exclusively found in the thin disk, the only Galactic population young enough to harbour intermediate-mass stars. The below-EHB population seems to be absent in other large samples, which generally include more thick disk and halo members. These findings suggest that non-degenerate formation channels may play a more prominent role in the Galactic disk than previously thought. 
}


   \maketitle

\section{Introduction}
Hot subdwarf stars are compact, typically helium-burning stars that have lost nearly all of their hydrogen envelopes.  
As a result, they are found on or near the extreme end of the horizontal branch \citep[EHB;][]{Greenstein_1974ApJS...28..157G,Newell_1973,Heber_1984} in the Hertzsprung--Russell Diagram (HRD), marking a transitional evolutionary phase between the first red giant branch (RGB) and the white dwarf cooling sequence. B-type hot subdwarfs (sdBs) are thought to be powered by core helium fusion and generally exhibit helium-poor atmospheres, with effective temperatures between 20\,000 and 40\,000~K and surface gravities ranging from $\sim$4.75 to 6.75 dex. In contrast, the more evolved O-type subdwarfs (sdO), which are assumed to be sustained by helium shell burning, show a broader range of helium abundances and temperatures extending from 40\,000\,K up to $\sim$80\,000\,K, along with slightly wider ranges of surface gravities and radii \citep[for comprehensive reviews, see][]{Heber_2009,Heber_2016_review,Heber_2026enap....2..488H}. 
Higher luminosity H-rich post-asymptotic giant branch (post-AGB) stars and central stars of planetary nebulae with surface gravities between roughly 4.5 and 5.5 dex are often spectroscopically classified as sdO as well, or as O(H) \citep{Reindl_2016A&A...587A.101R, Reindl_2024A&A...690A.366R_cspne}.

The formation of hot subdwarf stars remains an open question.
Since most sdBs have masses of about half a solar mass \citep[][]{Dorman_1993ApJ...419..596D_models, Fontaine_2012,Schaffenroth_2022_1}, close to the core-helium-flash mass, they are thought to descend from low-mass stars (0.7 to 1.8\,$M_\odot$) that have lost almost all of their H-envelopes while igniting helium under electron-degenerate conditions near the tip of the RGB. Such stars are typically referred to as canonical hot subdwarfs. However, intermediate mass (1.8 to 8\,$M_\odot$) progenitors can ignite helium under semi- or non-degenerate conditions (without a flash) in the subgiant or Hertzsprung gap phases. The resulting He-burning core masses can be as low as $\sim 0.33\,M_{\rm \odot}$ depending on the progenitor masses \citep{Han_2002, Hu_2008A&A...490..243H,Prada_2009JPhCS.172a2011P, 2018A&A...615A..78G, Arancibia_Rojas_2024MNRAS.52711184A}.

Although the predicted properties of He-burning cores closely resemble the observational properties of sdO/Bs, it is difficult to explain their extremely thin or completely missing hydrogen envelopes. The prevailing view is that binary interaction plays a crucial role in hot subdwarf formation \citep{Pelisoli_alone_2020}, following the evolutionary channels  detailed in \citet{Han_2002, Han_2003}. 
Binary population synthesis studies (BPS) suggest three main formation channels for hot subdwarfs. Stable Roche lobe overflow (RLOF) from a RGB star to a main-sequence (MS) companion of FGK-type can lead to hot subdwarfs in binaries with long orbital periods \citep[500 to 1500\,d;][]{Vos_2020A&A...641A.163V}. 
For low-mass companions, the mass transfer becomes unstable and leads to common envelope (CE) evolution, yielding hot subdwarfs with M-type MS, brown dwarf, or white dwarf companions on short-period orbits \citep[hours to days;][]{Schaffenroth_2022_1}. 
Finally, helium-rich sdO (He-sdO) stars are suggested to form through the late ignition of He-burning triggered by mergers involving He-core white dwarfs \citep{Webbink_1984ApJ...277..355W}. 
Related stars suggested to form through white dwarf mergers include the cooler and extreme helium (EHe) stars \citep{Jeffery_2020JApA...41...48J}, as well has the hotter O(He) stars \citep{Reindl_2014A&A...566A.116R_OHe}.
Single star evolution channels remain under discussion, such as internal mixing mechanisms in the hot flasher scenario \citep[see e.g. ][]{Sweigart_1997,Castellani_1993,Lanz_2004ApJ...602..342L,Marcello_2008A&A...491..253M, Battich_2018A&A...614A.136B}.

Current spectroscopic samples of hot subdwarfs, such as those based on the Sloan Digital Sky Survey \citep[SDSS;][]{Geier_2017, Kepler_2019, Geier_2024A&A...690A.368G}, the Large Sky Area Multi-Object Fiber Spectroscopic Telescope \citep[LAMOST;][]{Lei_2018ApJ...868...70L,Lei_2019ApJ...881..135L, Lei_2020ApJ...889..117L, Lei_2023ApJ...942..109L_spectra, Luo_2019ApJ...881....7L, Luo_2021}, or literature compilations \citep{Geier_2020,Culpan_2022} are affected by strong selection effects. 

With the advent of the \textit{Gaia} survey, all-sky volume-complete samples can be characterised in detail by combining spectroscopic and kinematic properties of the local stellar population.
In \citet{Dawson_2024A&A...686A..25D} (hereafter Paper I), we took the first step toward this goal by compiling a spectroscopically complete, volume-limited hot subdwarf sample to 500~pc, and presented classifications and a detailed discussion of space densities and spatial distributions. 
Our 500~pc sample, constructed using precise parallax measurements from \textit{Gaia}'s Early Data Release 3 \citep{Gaia_collaboration_2020yCat.1350....0G} in combination with the all-sky candidate catalogue of \citet{Culpan_2022}, provides a representative benchmark for the local population of hot subdwarfs. The stringent constraint on distance minimises selection biases. Furthermore, within 500~pc, all hot subdwarfs are sufficiently bright to be detected by \textit{Gaia}, resulting in an estimated completeness approaching 100\% for the single-lined, non-composite objects (see Paper I).

From our dataset, \citet{Dawson_2024A&A...686A..25D} derived a mid-plane space density of $\rho_{0}\,=\,6.03\pm 0.51 \times10^{-7}$ stars/pc$^{3}$, assuming a hyperbolic secant vertical profile, which has been widely adopted in previous studies \citep[e.g.][]{Gilmore_1983MNRAS.202.1025G,Villeneuve_1995a, Villeneuve_1995b,Bilir_a_2006NewA...12..234B,Yoachim_2006,Widrow_wave_discovery_2012ApJ...750L..41W,Xiangcheng_2017MNRAS.467.2430M,Canbay_2023AJ....165..163C}. This measured density is in strong tension with theoretical expectations: formation models predict space densities nearly an order of magnitude higher \citep{Han_2002, Han_2003, Clausen_2012ApJ...746..186C}, while even the most recent theoretical estimates still give densities at least four times larger than our result \citep{Nicolas_2_2025PASA...42...12R}.

In this paper, we performed a homogeneous atmospheric analysis of all identified hot subdwarf stars in our 500~pc sample. Following this, we focus our analysis on a restricted region within the \teff\ -- \logg\ parameter space where (post-)EHB stars are predicted to evolve, as described in Sect. \ref{sample_selection}.
Details about the spectroscopic observations are presented in Sect. \ref{further_observations}. In Sect. \ref{model_atmospheres} we describe the model atmospheres and the fitting procedure, which is followed by a detailed analysis of the systematic uncertainties. The kinematic methodology is described in Sect. \ref{kinematics}. All results are presented in Sect. \ref{results}, which are then discussed and compared with theoretical evolutionary tracks from \citet{Han_2002,Han_2003} in Sect. \ref{discussion}.

\section{Sample selection}
\label{sample_selection}
As mentioned above, the volume-complete sample presented here was compiled and fully spectroscopically classified in Paper I, which was directly drawn from \citet{Culpan_2022}, who selected candidates based on \textit{Gaia} colour and absolute magnitudes. 
For this paper, we started by selecting all 257 single-lined hot subdwarf targets from Paper I. 
Our analysis is focussed on a relaxed parameter range of 17\,000\,K $< T_\mathrm{eff}<$ 80\,000\,K and $4.50< \log g <6.75$, encompassing both pre- and post-EHB stars. Stars outside this range are beyond the scope of this paper and will be addressed in a separate study. 
This includes four well-known central stars of planetary nebulae with sdO-like cores: NGC 1360, PN\,A66\,36, KPD0005+5106, and MWP\,1, classed in the literature as O(H), O(H), O(He), and PG\,1159, respectively, which feature very high effective temperatures (\teff\ $\gtrsim 100\,000$\,K).

EC 11575-1845 displays an extreme reflection effect in its light curve \citep{Derekas_2015ApJ...808..179D}, which affects the modelling of its optical spectrum. We therefore removed this object from further analysis. 
PG\,1544$+$488 is an identified He-sdOB$+$He-sdB \citep{Jeffery_2014MNRAS.440.2676S} and is also removed from this paper and will be analysed as part of the binary population in a forthcoming paper.
We further exclude the 48 hot subdwarfs identified in Paper I (table A.2) that likely host MS companions of type A, F, G, or K; they will be discussed in a follow-up paper. This includes NGC 1514, a known O(H) star with a bright MS A-type companion.
Lastly, two stars in table A.3 from Paper I have now been found to be sdBs. This gives a total of 253 objects selected for an atmospheric analysis in our study.

\section{Spectroscopic observations}
\label{further_observations}

\subsection{Observational material}

\begin{table*}[!]
\setstretch{1.2}
\centering
\caption{Overview of the spectroscopic observations. }
\label{observation_log}
        \begin{tabular}{cccccccc}
                \toprule\toprule
                Telescope & Instrument&  Spectra &Set-up & Range & $\Delta\lambda$  & $\sigma_{v_\mathrm{rad}}$  & Programme ID \\
                &&&&[\AA]&[\AA]&[\kms]& \\
                \midrule
                INT & IDS/EEV10 & 1042 & R1200B & 3600 - 5100 & 1.0 & 5 & ING.NL.23B.005\\
         && &  &&&&ING.NL.23B.002\\  
         && &  &&&&ING.NL.24A.005\\
SOAR  & Goodman & 1008 & 930 M2 &3550 - 5300 & 3.2 & 15 & 2023B-711595\\
  &  &  &  & &  && 2024B-520098\\
  &  &  &  & &  && 2025A-537257\\
NOT  & ALFOSC & 339 & Grism 18 & 3450 - 5350 & 2.2 & 15 & 3-NOT1/25A \\
  & FIES & 57 & low-res & 3630 - 7170 & $R$=25\,000 & 2 & 3-NOT1/25A \\
Mercator  & HERMES & 188 & HRS &3800 - 9000 & $R$=85\,000 & 1 & GTO \\
NTT  & EFOSC2 & 40 & Grism 19 & 4441 - 5114 & 1.5 & 10 & 0114.D-0104(A) \\
WHT & ISIS & 1 & R600B & 3648 - 5067 & 1.76 & 10 & W13AN011\\
        \midrule
        Other & &  &&&&&Source\\ 
        \midrule
        MPG/ESO  & FEROS & 250 & - & 3500 - 9200 & $R$=48\,000 & 1 & ESO archive \\
        VLT UT3/ESO  & X-shooter & 227 & UVB/VIS & 3660 - 7100 & $R$=9861/18350 & 3& ESO archive \\
        LAMOST DR10 & - & 80 & LRS & 3690 - 9200 & 3.05& 15 &Database query\\
        VLT UT2/ESO  & UVES & 78 & various & various & various & 1 & ESO archive \\
        Mayall Telescope  & DESI & 4 & - & 3600 - 9800 & 2.09 & 15 & Database query \\
        \bottomrule
        \end{tabular}
\end{table*}

Paper I was based on one or more high signal-to-noise ratio (SNR > 100), low-resolution spectra ($R$ $\sim1200 - 2000$) for each unclassified candidate in the 500\,pc sample for classification. 
For this paper, in addition to the spectra obtained for paper I which are also analysed here, we obtained high-SNR spectra for sources where their data were inaccessible in the literature to enable a homogeneous quantitative spectroscopic analysis for all stars. 

In addition, we include multi-epoch medium- to high-resolution spectra obtained in an ongoing campaign for all hot subdwarf stars within the 500 pc sample (Dawson et al. in prep). 
These spectra help to assess binarity and allow us to estimate true space velocities for a kinematic analysis of the sample\footnote{A detailed study of the binary population will be given in a follow-up paper.} (see Sect.\ \ref{kinematics}).  
Further observations began in August 2023 and mainly used the Intermediate Dispersion Spectrograph at the Isaac Newton telescope (INT/IDS), the Goodman spectrograph at the Southern Astrophysical Research Facility (SOAR/Goodman), and the Alhambra Faint Object Spectrograph and Camera at the Nordic Optical Telescope (NOT/ALFOSC), covering the optical range from approximately 3600\,\AA\ to 5200\,\AA. Using a $0.5^{\prime \prime}$ or $1.0^{\prime \prime}$ slit, each setup achieves spectral resolutions of 1\,\AA, 2.2\,\AA, and 2.6\,\AA, respectively, corresponding to radial velocity ($v_\mathrm{rad}$) accuracy ($\sigma_{v_\mathrm{rad}}$) between 5 and 15 \kms where the lower dispersions give rise to higher inaccuracies in the wavelength calibration.
We are currently also obtaining additional spectra using the High-Efficiency and High-Resolution Mercator Echelle Spectrograph \citep[HERMES;][]{hermes_2011A&A...526A..69R} on the 1.2 meter Mercator telescope and on the FIbre-fed Echelle Spectrograph (FIES) at the NOT in the northern hemisphere. The HERMES instrument offers a high spectral resolution of $R =85\,000$, while FIES provides $R=25\,000$ in its low-resolution mode. These resolutions correspond to radial velocity precisions ($\sigma_{v_\mathrm{rad}}$) of less than $1$ \kms\ and $2$ \kms, \,respectively, limited by the broad and relatively sparse hydrogen Balmer and helium lines present in hot subdwarf stars.

We considered public spectroscopic surveys for our analysis. 
This includes the LAMOST DR10 \citep[][]{LAMOST_2022yCat.5156....0L} low- and medium-resolution spectroscopic surveys (LRS/MRS) and the first data release of the Dark Energy Spectroscopic Instrument \citep[DESI;][]{desicollaboration2025datarelease1dark}. 
The European Southern Observatory (ESO) archive\footnote{http://archive.eso.org/eso/eso\_archive\_main.html} included further spectra from the X-shooter, FEROS, and UVES instruments. None of our targets appear in the SDSS archives, as they are brighter than the survey's upper magnitude limit.
A summary of all spectra analysed is provided in Table \ref{observation_log} with some examples presented in Fig. \ref{sdB_spectrum}.

In total, our spectroscopic dataset study includes 3226 spectra for 253 single-lined hot subdwarf stars where each source has at least three spectra suitable for both atmospheric analysis and \vrad\ determination.
The spectroscopic data were reduced either using PyRAF procedures \citep{PYRAF_2012ascl.soft07011S}, a python-based implementation of IRAF \citep[Image Reduction and Analysis Facility;][]{IRAF_1986SPIE..627..733T} which is a general purpose software developed by the National Optical Astronomy Observatories (NOAO), the \textsc{Molly} package \citep{Marsh1989optimalExtraction,Marsh2019Molly}, or instrument-specific pipelines. All methods include basic bias and flat-field corrections and wavelength calibrations.

\subsection{Revised spectroscopic classifications}
\label{classification}
The classification scheme used in Paper I was based on the scheme outlined in \citet{Moehler_1990A}, which was updated and extended to incorporate a wider range of spectroscopic classes (see Paper I for details). It was based on visual inspection and is adopted in this work. In addition to the basic spectral classes of sdB, sdOB, sdO, and He-sdO, here we also incorporate intermediate helium-sdOBs (iHe-sdOB), and extreme helium-sdOs (eHe-sdO) into our scheme, as already adopted in other works \citep{Luo_2021, Dorsch_2024PhDT........36D, Geier_2024A&A...690A.368G}. Visually, both iHe-sdOB and eHe-sdO stars exhibit strong neutral and ionized helium lines. However, eHe-sdO stars are characterized by weak or even absent hydrogen lines. 
We classify these subgroups based on their helium abundances derived from spectral fitting. We define iHe-sdOB stars as those with helium abundances between \logy\ = $-$1.2 and \logy\ = 0.6, while stars with \logy\ > 0.6 are classified as eHe-sdO. The divisional point of the helium-enriched stars is motivated by a clear gap that is seen in the \logy\ - \teff\ parameter space at \logy\ = $-$1.2 (e.g. see Fig. \ref{helium}). 
Lastly, we point out two misclassified systems in Paper I: [L92b]MarkA and TYC6017-419-1. These were classified as sdO and sdOB respectively. They have, however, been reassigned as eHe-sdOs herein. Table \ref{main_table_atm} includes the most accurate and current information.

\section{Atmospheric and stellar parameter determination}
\label{model_atmospheres}

\subsection{Model atmospheres and synthetic spectra}

\begin{figure*}[h!]
    \centering
    \includegraphics[width=1\linewidth]{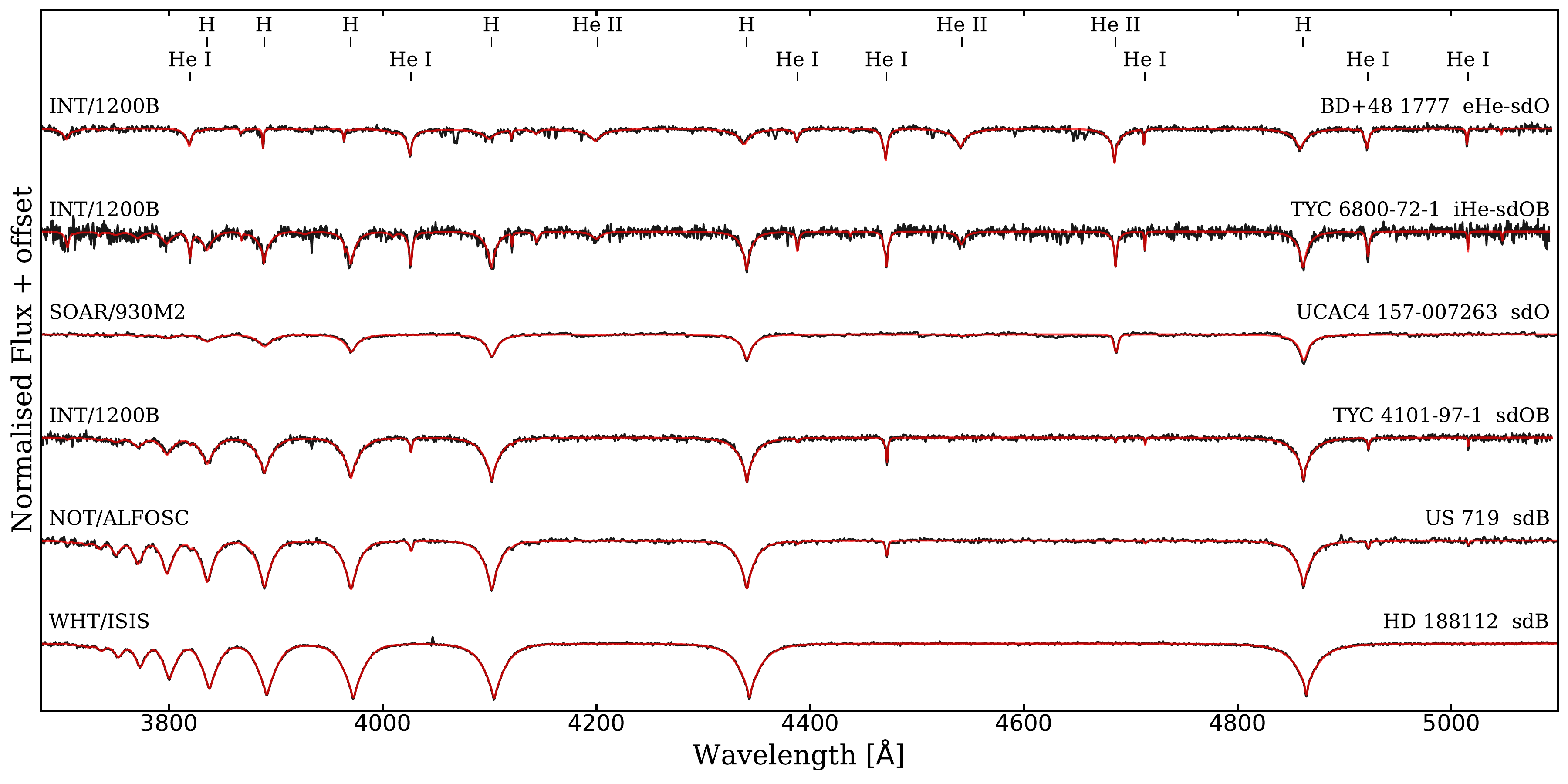}
    \caption{Example spectral fits (red) to various hot subdwarfs (black) using the pipeline described in Sect.\ \ref{model_atmospheres}. 
    Table \ref{observation_log} states more details on the spectra. 
    }
    \label{sdB_spectrum}
\end{figure*}

For the determination of the atmospheric parameters effective temperature (\teff), surface gravity (log $g$) and helium abundance (\logy), we applied the same state-of-the-art set of model atmosphere grids (the second generation Bamberg model grids) to all hot subdwarf stars for a homogenous analysis \citep[also used in][]{Heber2025, latour2025arizonamontrealspectroscopicsurveyhot, Geier_2024A&A...690A.368G}, which is detailed below. For each stellar spectrum we performed a full spectral fit. Spectra that did not cover the hydrogen Balmer jump (covering at least about 3700\,\AA\ to 5000\,\AA) were not considered in the atmospheric parameter determination, and only used for the radial velocity calculation. 
The same model grid was used for all stars, spanning $9000$ to $75\,000$\,K in \teff, $3.8$ to $7.0$ in \logg, and $-5.05$ to $+2.64$ in \logy; regions beyond the Eddington limit (low \logg\ at high \teff) are excluded. A hybrid non-local thermodynamic equilibrium (NLTE) approach was adopted to balance computational efficiency with the need to capture NLTE effects in hot early-type stars. Synthetic spectra were generated with the \atlas, \detail, and \surface\ (ADS) codes, as in previous hot subdwarf studies \citep{Latour_2018A&A...618A..15L, Schneider_2018, Geier_2024A&A...690A.368G}. Specifically, \atlas\ \citep{Kurucz_1996} computed the LTE temperature and (electron) density stratifications; \detail\ \citep{Giddings_1980PhDT.......131G} solved the radiative transfer and rate equations for hydrogen and helium populations, iteratively feeding corrections back into \atlas\ \citep{Irrgang2018}; and \surface\ \citep{Giddings_1980PhDT.......131G} produced the final synthetic spectra including line broadening.

The model fits were carried out using the Interactive Spectral Interpretation System \citep[ISIS;][]{ISIS_2000ASPC..216..591H}, employing a $\chi^2$-minimisation technique to perform a global fit across the entire usable spectral range. A detailed description of the implementation of spectral fits in ISIS can be found in \citet{Irrgang_2014}. The routine was automated such that the fits could be done in parallel, and takes about two hours to complete in total for our sample of 253 single-lined hot subdwarfs on a multi-core cluster. 

In the automated spectroscopic fitting procedure, the continuum is modelled with an Akima spline \citep{Akima1970}, which is anchored at $100\,\AA$ intervals while avoiding the hydrogen and helium absorption features in the spectrum. The atmospheric parameters \teff, \logg, and He-abundance are fitted simultaneously with the continuum, in addition to the radial velocity. Poorly fitting spectral regions are automatically excluded. For targets with multiple spectroscopic observations, each individual spectrum was fit separately. This method is the same as that used in \citep{Heber2025} and \citet{latour2025arizonamontrealspectroscopicsurveyhot}.

For the atmospheric parameters \teff, \logg, and \logy, and their associated statistical uncertainties, we adopted weighted averages, where each measurement was weighted by the square of the effective signal-to-noise ratio (SNR\textsubscript{eff}) of the corresponding spectrum -- that is, an inverse-variance weighting. The SNR\textsubscript{eff} values were computed after rescaling the flux uncertainties such that the reduced $\chi^{2} = 1$ for each spectral fit. This approach assigns greater weight to higher-quality spectra while preventing spectra with more data points from being overrepresented in the combined values. 

The spectral resolution for each fit is set by the average full-width at half-maximum (FWHM) of the arc lamp emission lines, which reflects the instrumental broadening of the spectrograph, or taken from archival values when available. Each fit was then visually inspected to assess its quality.
Spectra with SNR $<$ 5 were discarded. In cases where poor wavelength calibration was detected during visual inspection, efforts were made to reprocess the data to obtain a usable spectrum; if re-reduction was unsuccessful, the spectrum was excluded from the fitting routine.

The routine was performed twice: first with the projected rotation \vsini\ constrained to values below 30 \kms\ typical for hot subdwarf stars \citep{Geier_heber_2012A&A...543A.149G} and then with \vsini\ treated as a free parameter to enable the detection of rare, rapidly rotating hot subdwarfs. No such candidates were identified using the high-resolution data. 
In total, the routine produced 3226 reliable spectroscopic fits after 88 were discarded due to poor quality. The results with \vsini\ constrained below 30 \kms\ were used to ensure reliable \logg \, estimates, thus avoiding the correlation between \vsini\ and \logg.

\subsection{Systematic uncertainties}
\label{systematic_uncertainties}

\begin{figure*}[h!]
    \centering
    \begin{subfigure}[b]{0.32\linewidth}
        \centering
        \includegraphics[width=\linewidth]{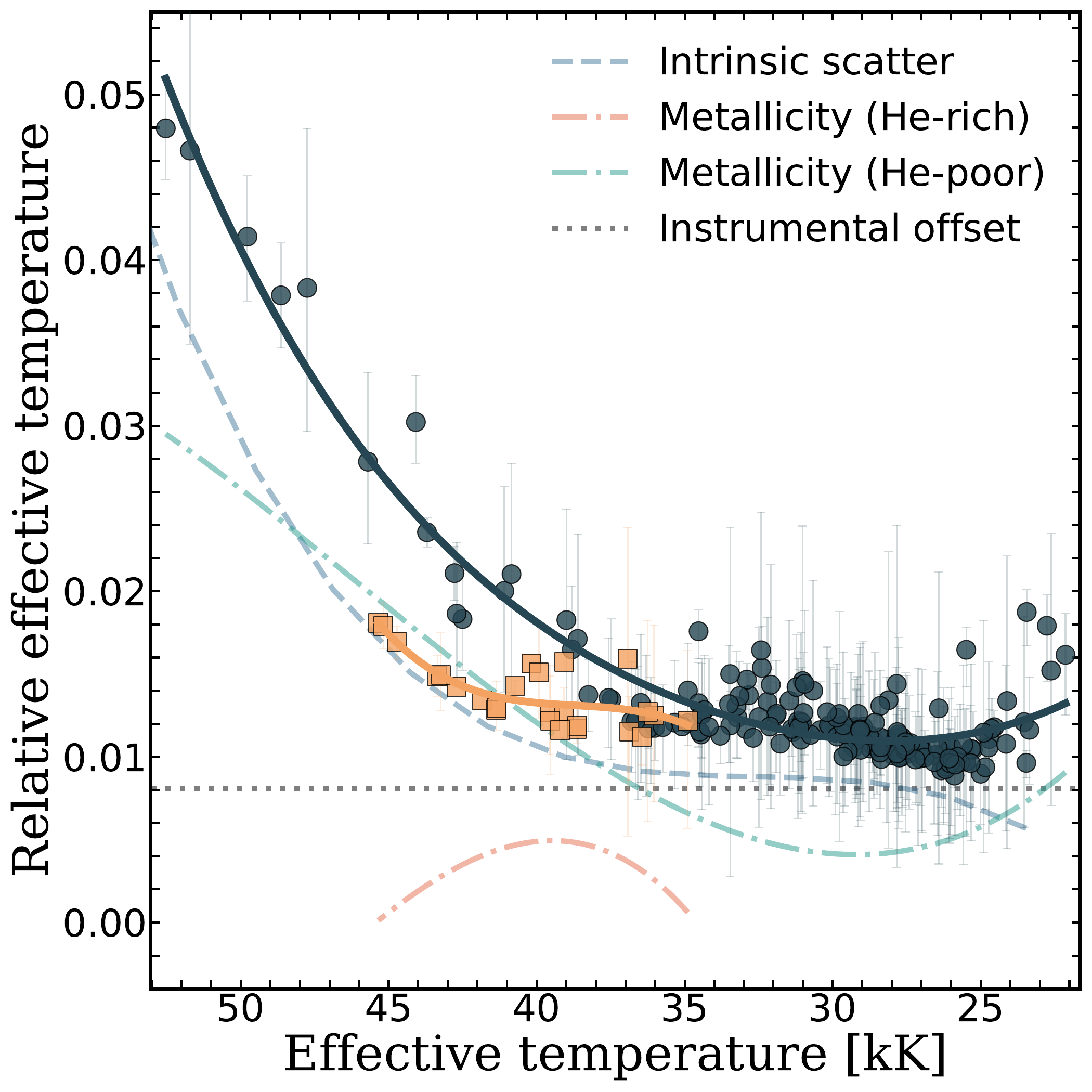}
    \end{subfigure}
    \hspace{0.001\linewidth}
    \begin{subfigure}[b]{0.32\linewidth}
        \centering
        \includegraphics[width=\linewidth]{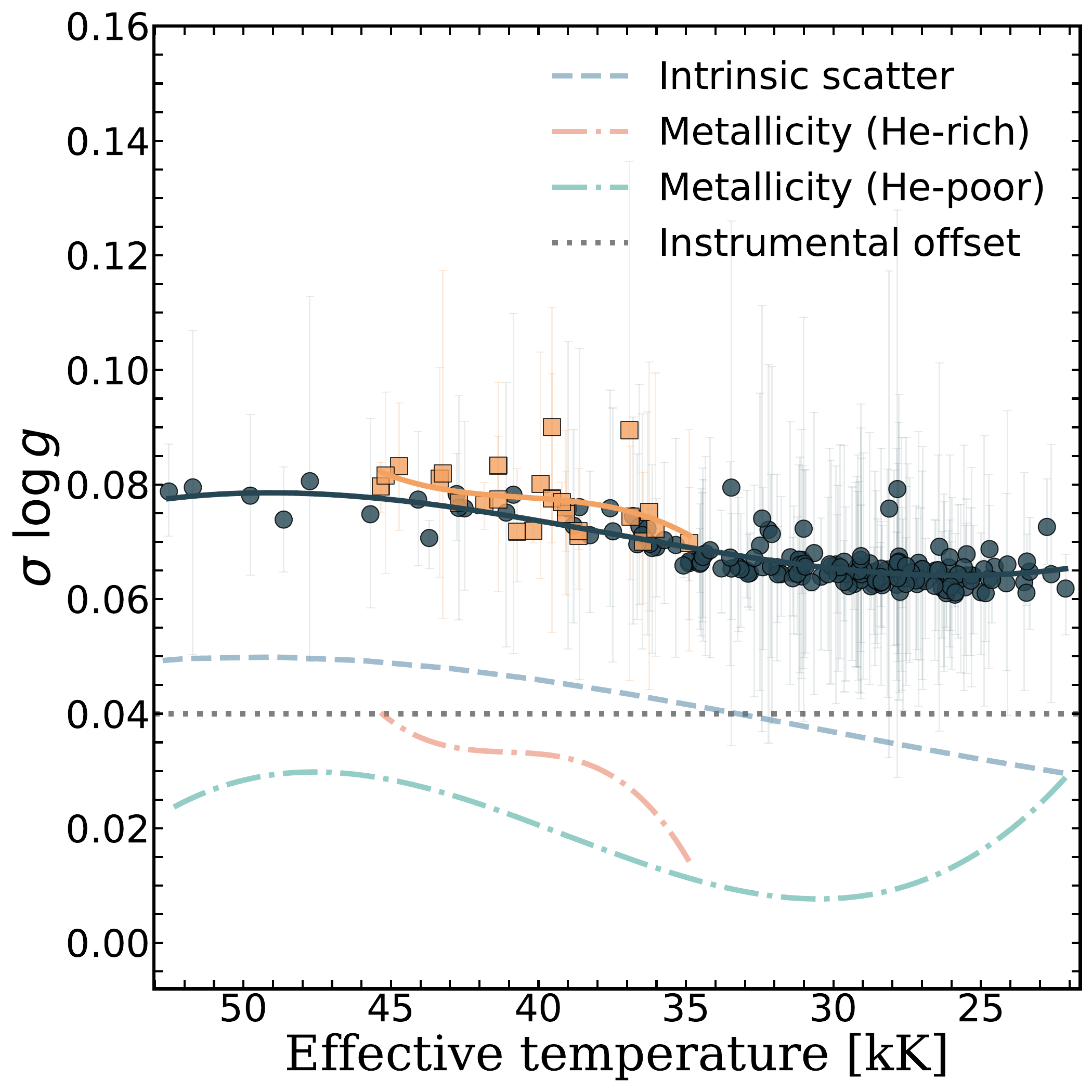}
    \end{subfigure}
    \hspace{0.001\linewidth}
    \begin{subfigure}[b]{0.32\linewidth}
        \centering
        \includegraphics[width=\linewidth]{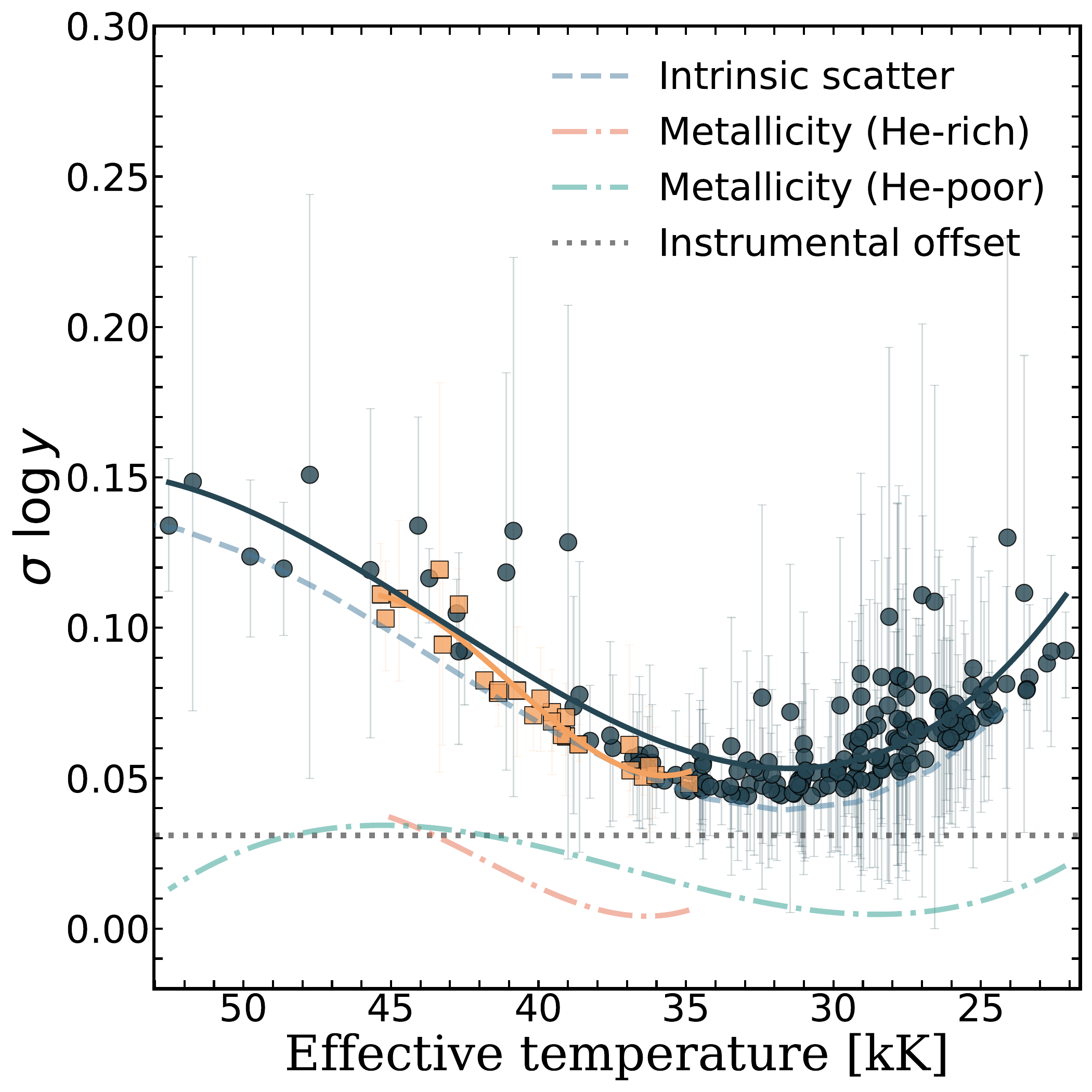}
    \end{subfigure}
    \caption{Three-panel figure showing our best estimates of the overall uncertainties in \teff, \logg, and \logy\ (from left to right), displayed as scattered markers. \teff\ uncertainties are plotted as fractional values ($\Delta T_{\mathrm{eff}}/T_{\mathrm{eff}}$, i.e. 1–5\%). Helium-rich and helium-poor stars are separated and indicated as orange squares and dark blue circles, respectively. The solid orange and dark blue lines are polynomial fits (equation \ref{uncertainty_polynomial}) to these scattered markers, and the coefficients of these are given in Table \ref{tab:uncertainty_coeffs}. The dashed polynomials represent the identified uncertainty contributions from the intrinsic scatter measured in our INT data and from the metallicity abundance uncertainty, as described in Sect. \ref{systematic_uncertainties}. The formal statistical errors are given by the error bars of the scattered markers and the instrumental offset between INT and ALFOSC is given by a black-dashed line. All four identified uncertainties are added together in quadrature to arrive at the uncertainties used for the analysis in this work.}
    \label{systematic_uncertainty_final}
\end{figure*}

Our automated and homogenous spectral analysis approach is intended to minimise internal systematic uncertainties that would be introduced by the use of multiple model grids or fit methods. However, other aspects of data handling contribute to further uncertainty. 
Next to the formal uncertainty caused by statistical noise, we identify and characterise systematic uncertainties that arise from three main sources:
(1) variability introduced by data reduction and observing conditions (e.g.\ clouds),
(2) calibration offsets between different instruments and observing setups,
and (3) the choice of metallicity used in the spectral fitting procedure.
In the following we detail these systematic uncertainties.

Since our spectral fits were performed independently for each spectrum, we can directly assess the repeatability of parameter estimates under similar conditions. The majority of our spectra were obtained with the INT/IDS instrument using the 1200B setup (see Table \ref{observation_log}), which covers the full optical range except for H$\alpha$. For each star with at least three INT spectra at SNR\textsubscript{eff} > 25, we computed the one-sigma observed scatter in \teff, \logg, and \logy\ around their respective weighted means. 
By comparing this observed scatter to that expected from formal uncertainties, we isolate the intrinsic scatter likely associated with data reduction and observing conditions. 
This contribution amounts to, on average, 0.55\,\% in \teff, 0.039 in \logg, and 0.02 in \logy. More precisely, the contribution is a function of \teff\ as presented by a light dashed-blue polynomial in each panel of Fig. \ref{systematic_uncertainty_final}, and labelled as the intrinsic scatter.

The largest set of overlapping observations between different instruments occurred for nine stars observed with both INT and ALFOSC (150 spectra between them in total). To estimate systematic offsets in \teff, \logg, and \logy\ between the two setups, we first computed uncertainty-weighted mean values and associated errors for each star from each instrument. The per-star differences were then modelled using a maximum likelihood approach that fits simultaneously for a constant systematic offset and an additional intrinsic scatter term beyond the reported statistical uncertainties. This method provides an unbiased estimate of the systematic offset while accounting for excess variance not captured by formal errors. The resulting absolute systematic offsets are $0.81 \pm 0.13$\,\% in \teff, $0.04 \pm 0.01$ in \logg\ and $0.03 \pm 0.01$ in \logy, which are global offsets which we take as a proxy to represent the uncertainties that arise from using different instruments. These are given as horizontal dotted-black lines in Fig. \ref{systematic_uncertainty_final} and are labelled as the instrumental offset. It is important to emphasise, however, that adopting a single offset derived from INT and ALFOSC as representative of all instrumental pairings in our program is a strong simplification. Intrinsic differences between spectrographs cannot, in general, be encapsulated by a universal correction. We underscore that our adopted offsets are tailored to this specific comparison, and any assumed cross-instrument calibration should be evaluated with care for the instruments and samples at hand.

All atmospheric parameters presented in this study are based on fits assuming standard sdB atmospheric metallicity \citep[$\log z/z_\mathrm{sdB}=0$; using abundances from][]{Pereira2011}. Accurate measurements of atmospheric metallicity typically require high-quality ultraviolet (UV) data, which is available for only a limited number of hot subdwarf stars. Due to the opacity from numerous iron-group elements in the far-UV to extreme-UV range, the metallicity significantly affects the derived atmospheric parameters.  
The iron abundance serves as an indicator of this metallicity; it scatters with a standard deviation of $0.3$ dex in the sample of \citet{Geier_2013A&A...549A.110G}. 
To account for this, we repeated the entire fitting routine for all stars with fixed metallicities of $\log z/z_\mathrm{sdB}=+0.3$ and $\log z/z_\mathrm{sdB}=-0.3$ \citep[see also][]{Heber2025}. 
This variation leads to systematic uncertainties as a function of \teff, given by the dot-dashed polynomials in Fig. \ref{systematic_uncertainty_final}. In this case, since we fit all stars in the sample and have ample number statistics, we separate the stars into helium-poor and helium-rich objects, which returns a distinct behaviour for each group. For further visualisation of the impact of metallicity in the \logg\ - \teff\ and \logy\ - \teff\ parameter spaces, see Fig. \ref{fig:metallicity}.

To derive the final systematic uncertainties on \teff, \logg, and \logy, we combined the three isolated systematic uncertainties detailed above in quadrature, which results in the scatter markers in each panel of Fig. \ref{systematic_uncertainty_final}.
To the total uncertainties in Fig. \ref{systematic_uncertainty_final}, we fit a third-order polynomial of the form
\begin{equation}
\label{uncertainty_polynomial}
y = a_3 x^3 + a_2 x^2 + a_1 x + a_0
\end{equation}
where the best-fit coefficients for each parameter \teff, \logg, and \logy\ for both helium-poor and helium-rich objects are provided in Table \ref{tab:uncertainty_coeffs}. Note that the formal statistical uncertainties are not included in these fits but are shown separately as error bars, allowing our estimates of the systematic uncertainties to be used independently in other works. In this paper, however, we combine the statistical uncertainties with the three systematic uncertainties described above in quadrature, and use these combined values for the remainder of the analysis.

These represent our best estimates of the total uncertainty in the atmospheric parameters, incorporating both random and systematic sources of error.

\subsection{Spectral energy distributions and stellar parameters}
\label{sed_fitting}
For the single-lined hot subdwarf stars in our sample, we applied the spectral energy distribution (SED) fitting routine first described in \citet{Heber2018}, and in more detail in section 2.2 of \citet{Dorsch_2024PhDT........36D}. This method compares apparent multi-band magnitudes collected from the literature though the VizieR database\footnote{https://vizier.cds.unistra.fr/} with synthetic SEDs, while keeping the stellar parameters fixed to the spectroscopically determined values. We refer the reader to \citet{Culpan_2024A&A...685A.134C} for a list of all the photometric catalogues queried. Interstellar extinction was modelled using the prescription of \citet{Fitzpatrick_2019ApJ...886..108F}, adopting a standard extinction coefficient of $R(55)=3.02$. 
Inspection of the SED further allows us to detect the presence of an IR-excess, which is visible if a cool MS companion is bright enough to emit more IR flux than the primary hot subdwarf star. While Paper I identified 48 out of the 305 hot subdwarfs in the 500~pc sample to likely harbour F, G, or K-type MS companions, a re-analysis of the SED fits identified a further nine of the 253 single-lined systems as showing a hint of IR excess, indicative of a cool companion star (see Table \ref{IR_excess} for a list of these objects). 
Because the small amount of IR flux in these systems has little to no effect on the derived atmospheric parameters from the optical spectra, they are included in our spectroscopic analysis; other composite-colour binaries were excluded. In the case of the SED fits, the flux contribution of the companion was modelled using the \textsc{Phoenix} Göttingen spectral library \citep{Husser_2013}\footnote{http://phoenix.astro.physik.uni-goettingen.de/}. 
All SED fits were constrained using the atmospheric parameters from our spectroscopic fits (\teff, \logg, and He-abundance) along with their corresponding systematic uncertainties (Sect. \ref{systematic_uncertainties}). Thanks to the precise parallax measurements from the \textit{Gaia} space mission \citep{Gaia_collaboration_2023A&A...674A...1G}, the interstellar reddening can be constrained, which allows absolute stellar radii $R$, luminosities $L$, and masses $M$ to be determined with high accuracy and minimal systematic uncertainties, particularly for our sample of nearby objects, which are evaluated using the following relations:

\begin{equation}
    R = \frac{\theta}{2 {\varpi}}, \quad
    M = \frac{g R^2}{G}, \quad
    L = 4\pi R^2 \sigma_{\mathrm{SB}} T_{\mathrm{eff}}^{4}, \quad
    \label{sed_params}
\end{equation}

\noindent
where $\theta$ is the angular diameter, $\varpi$ is the parallax, and $G$ is the gravitational constant. 

We employed a Monte Carlo method to assess the uncertainties in our calculations. 
The input spectroscopic parameters, \textit{Gaia} parallax, and angular diameter were represented by Gaussian distributions. 
For each star, $10^6$ samples were generated, with each array carried through the full calculation to the final derived parameter. 
The masses and luminosities listed in Table \ref{main_table_sed} are the median values of these distributions, while the asymmetric uncertainties are defined by the 84th and 16th percentiles, corresponding to the 68\% confidence interval around the median mass. 
We refer to these masses and luminosities, which can be constrained primarily due to the \textit{Gaia} parallax, as parallax-based masses and parallax-based luminosities for simplicity. 

\section{Kinematical method and population assignment}
\label{kinematics}

Hot subdwarf stars are found across all Galactic stellar populations. For less evolved stars, chemical tagging provides a powerful means of identifying stellar population membership \citep[see][for a full review]{Bland-Hawthorn_2016ARA&A..54..529B}. In hot subdwarf stars, however, strong diffusive processes in their atmospheres significantly modify the abundance patterns, rendering chemical tagging ineffective \citep{Michaud_2011A&A...529A..60M}. As a result, population studies of hot subdwarfs must instead rely on kinematical analysis, which requires complete 6D phase-space information \citep{Pauli2006A&A...447..173P}. 
From \textit{Gaia} we have accurate positions, parallaxes, and proper motions for all stars in our sample. 
The final parameter, $v_\mathrm{rad}$, is straightforward in principle but observationally intensive, especially given that a large fraction of hot subdwarfs that reside in close binaries exhibit velocity variations of hundreds of \kms\ and their orbital motion needs to be corrected for to obtain the radial component of the space velocity \citep{Maxted_2001,Napiwotzki_SPY_2004Ap&SS.291..321N,Copperwheat_2011MNRAS.415.1381C,Geier_2022,Schaffenroth_2022_1}

\subsection{Radial velocities}
All available spectra were processed with our automated fitting pipeline (Sect.~ \ref{model_atmospheres}) where the $v_\mathrm{rad}$ is an output of the full model fit. These are then corrected to the heliocentric frame.
For known close binaries, we adopt systemic velocities from the literature. For $\sim$30 systems solved in this work, we use systemic velocities derived from orbital solutions (Dawson et al. in prep.). For apparently non-variable or unsolved systems, we take the mean $v_\mathrm{rad}$ of the maximum and minimum $v_\mathrm{rad}$ values from the fitted spectra. This was chosen because of the poor orbital coverage for some binary systems. The standard deviation of these measurements is used as the uncertainty on the systemic radial velocity. 
Finally, systematic uncertainties were estimated for each instrumental setup and are listed in Table~\ref{observation_log}. Where available, we used the root mean square of the wavelength calibration derived during data reduction. For setups lacking direct calibration data, we estimated systematic uncertainties from the typical deviations of radial velocities measured for $v_\mathrm{rad}$-standard stars throughout the observing program. These systematic contributions were then added in quadrature to the statistical uncertainties to obtain the final error estimates.

\subsection{Kinematics}
\label{method:kinematics}

We compute the Galactic rest frame velocity components $U$, $V$, and $W$, defined as positive toward the Galactic Centre, the direction of Galactic rotation, and the north Galactic pole, respectively. These velocities are derived directly from each star's $v_\mathrm{rad}$, distance, position, and proper motion as provided by \textit{Gaia}, without assuming a specific Galactic potential or computing stellar orbits \citep[see][for details]{Johnson_Soderblom_1987AJ.....93..864J}. Distances were calculated by inverting the zero-point–corrected \textit{Gaia} parallaxes, with the parallax uncertainties inflated following \citet{ElBadry_2021}.
Velocities are transformed into the Galactocentric frame, adopting a solar distance from the Galactic centre $r_{\odot}=8.4$ kpc and a circular velocity of the local standard of rest (LSR) of $V_{\mathrm{LSR}}\,=\,242$\,\kms\ following the best-fit model of \citet{Irrgang_2013AA...549A.137I}.
The solar velocities with respect to the LSR were assumed to be $(U_{\odot}, V_{\odot}, W_{\odot}) = (11.1, 12.2, 7.3)$\,\kms\ based on the results of \citet{schoenrich_2010MNRAS.403.1829S}.  

Population membership is assigned probabilistically in $UVW$ space using a Gaussian mixture model (GMM) with three fixed components (thin disk, thick disk, halo). The velocity ellipsoid centroids, dispersions, and correlations are adopted from  \citet{Anguiano_2020}, while only the relative weights of the components are optimised using an expectation-maximisation (EM) algorithm. The standard EM-GMM framework \citep{Dempster1977, McLachlanPeel2000} provides a probabilistic decomposition of the data into a set of Gaussian components, but assumes error-free measurements, treating all observed scatter as intrinsic population variance. Our approach explicitly incorporates the per-object covariance matrix of the measurement errors into both the expectation and maximisation steps, following the extreme deconvolution methodology \citep{Bovy2011}, which has been widely applied to astronomical data with heteroscedastic uncertainties \citep{Hogg2010, Kelly2007}.

\begin{figure*}
    \centering
    \begin{subfigure}[b]{0.49\linewidth}
        \centering
        \includegraphics[width=\linewidth]{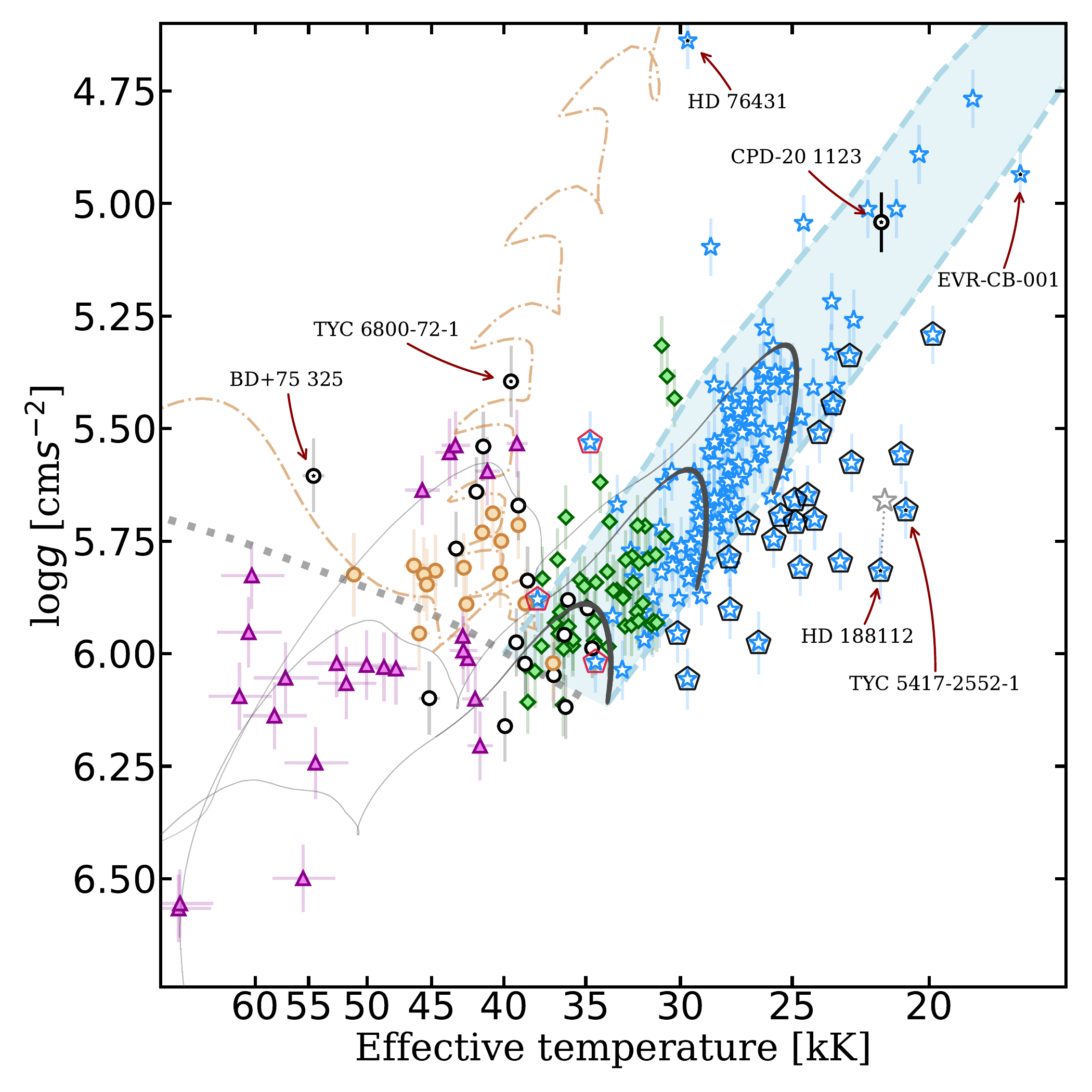}
        \caption{Effective temperature versus \logg\ (Kiel) diagram of the sample.}
        \label{kiel}
    \end{subfigure}
    \hfill
    \begin{subfigure}[b]{0.49\linewidth}
        \centering
        \includegraphics[width=\linewidth]{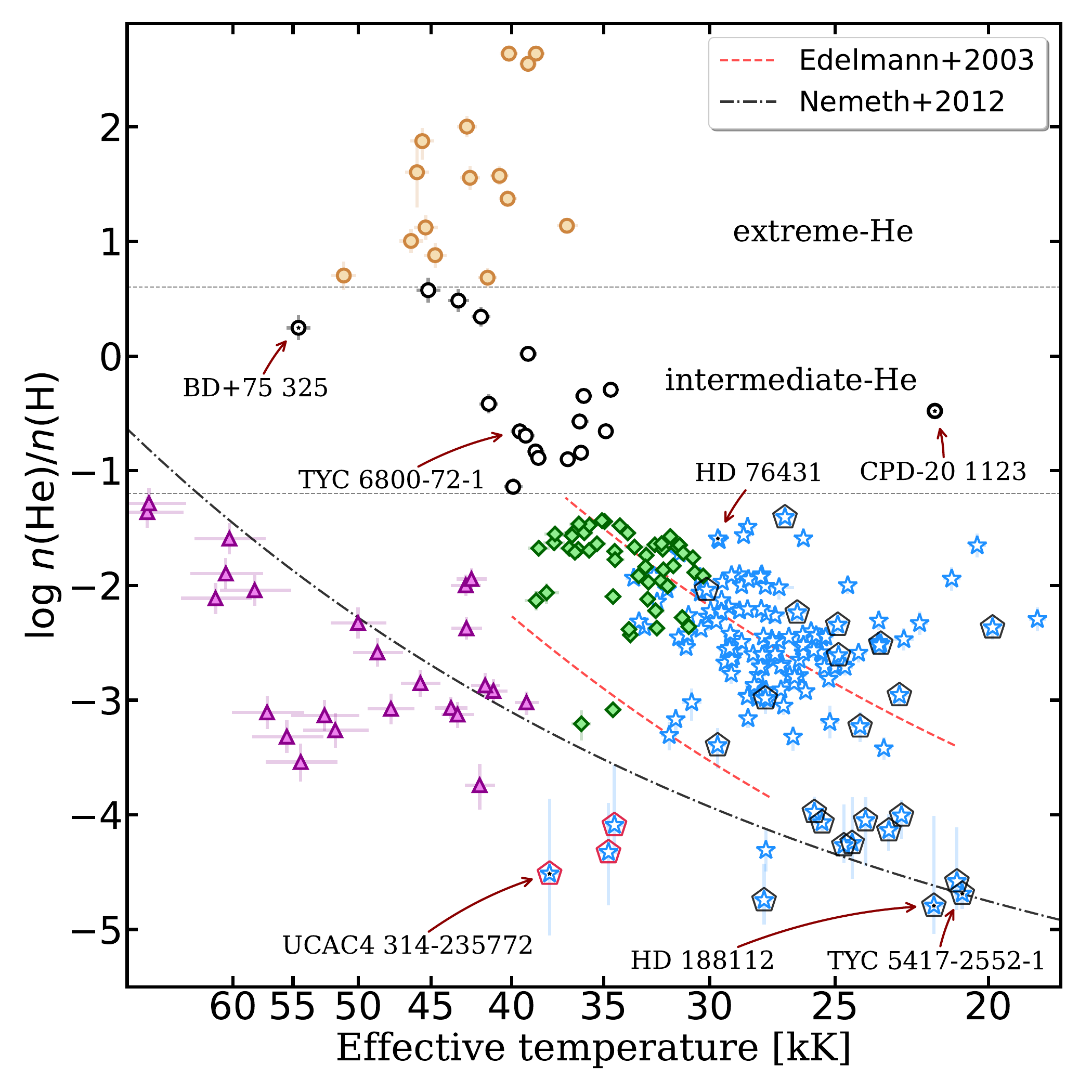}
        \caption{Helium abundance versus \teff.}
        \label{helium}
    \end{subfigure}
    \vspace{0.1cm} 
    \begin{subfigure}[b]{\linewidth}
        \centering
        \includegraphics[width=0.77\linewidth]{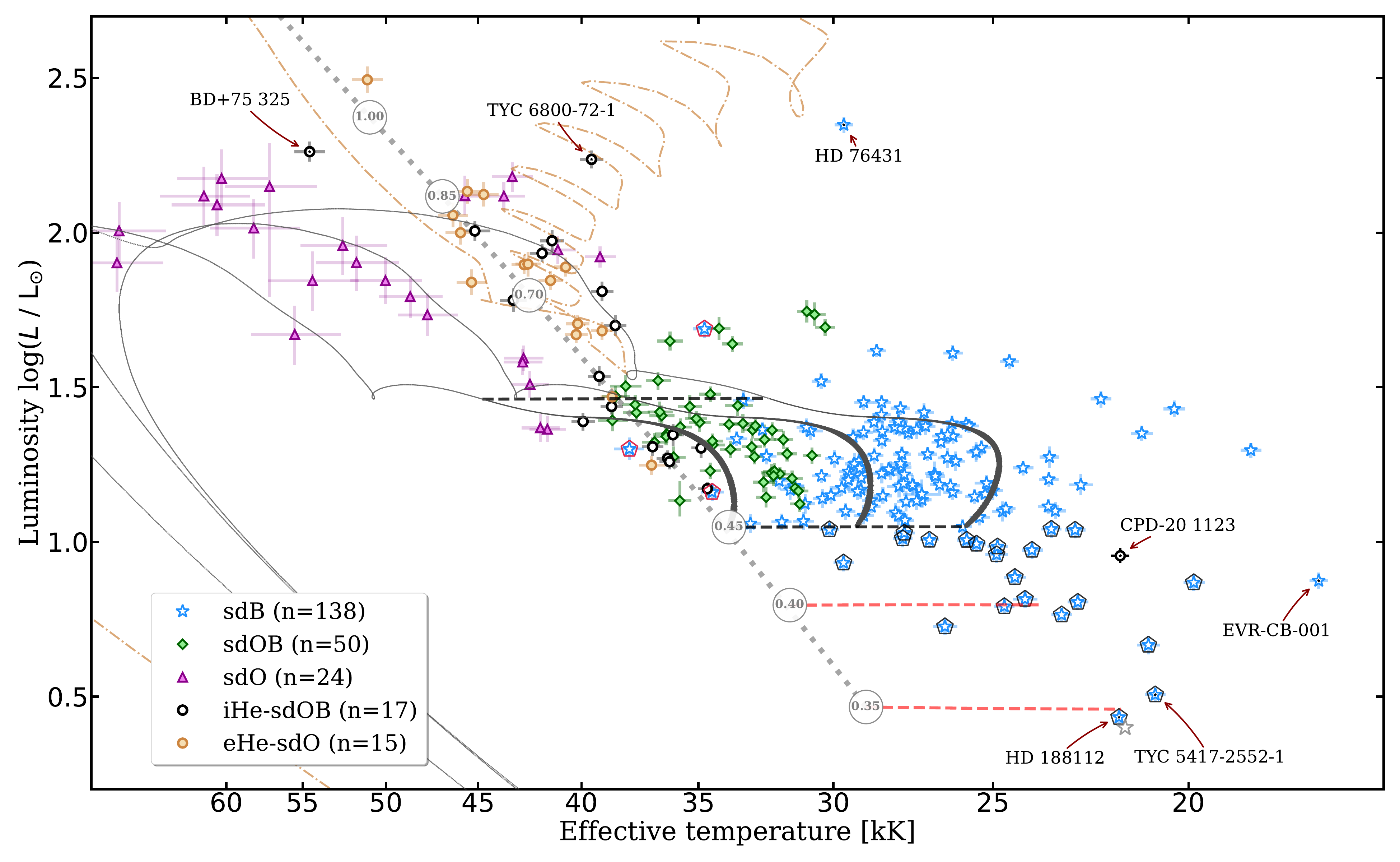}
        \caption{Hertzsprung-Russell diagram. Class counts are provided in the legend. }  
        \label{hrd}
    \end{subfigure}
    \caption{Upper left: $T_{\mathrm{eff}}$ - \logg\ diagram of the hot subdwarf stars in the 500~pc sample, colour-coded as shown in the legend (lower panel). Three post-EHB evolutionary tracks for three different envelope masses from \cite{Han_2003} are given as solid-black lines: 0.00, 0.001, and 0.005 \msun\ from bottom to top ($M_\mathrm{total} = 0.45\,M_{\odot}$). The thickness of the lines are linearly proportional to the predicted evolutionary time. A double He-WD merger track from \cite{Zhang_Jeffery_2012MNRAS.419..452Z} is shown as a orange dot-dashed line.
    The region encapsulated by the zero-age EHB and the terminal-age EHB from \citet{Dorman_1993ApJ...419..596D_models} is depicted as a shaded light-blue region which extends down to the He-MS from \citet{Paczynski1971} which is shown as a dotted-grey line. All tracks are given for solar metallicity.
    Upper right: helium abundance - $T_{\mathrm{eff}}$ diagram for the same stars. The helium sequences identified in earlier studies \citep{Edelmann_2003A&A...400..939E, Nemeth_2012MNRAS.427.2180N, Luo_2016, Lei_2018ApJ...868...70L} are shown.
    Bottom: Hertzsprung-Russell diagram with the same hot subdwarf classes and evolutionary tracks as shown in the upper left panel. The He-MS from \citet{Paczynski1971} is again shown as a dotted-grey line with the corresponding masses annotated. The horizontal dashed-black lines indicate the solar metallicity ZAEHB at $\log L/L_{\odot} = 1.05$ at the start of the $0.45$ \msun\ evolutionary tracks from \citet{Han_2002}, and the TAEHB of these tracks at the point of core helium depletion. Two additional ZAEHB positions are marked as dashed-red lines for the $0.40$ \msun\ and $0.35$ \msun\ evolutionary tracks to compare with the low-luminosity objects. Black pentagons mark the underluminous hot subdwarfs ($\log L/L_{\odot} < 1.05$) in all three plots. Crimson pentagons indicate the three identified hot and helium-poor stars. The grey star marks the previous position of HD\,188112.}
    \label{kiel_helium_hrd}
\end{figure*}

\section{Results}
\label{results}
Tables~\ref{main_table_atm}, \ref{main_table_sed}, and \ref{main_table_kin} summarise the results of our spectroscopic, photometric, and kinematic analyses, respectively.
Table~\ref{main_table_atm} lists the atmospheric parameters derived from our spectroscopic fits, including \teff, \logg, helium abundance, and systemic radial velocity. The stated uncertainties on the atmospheric parameters correspond to the systematic uncertainties derived in Sect.~\ref{systematic_uncertainties}, while those on the systemic radial velocities also include the $\sigma_{v_\mathrm{rad}}$ values provided in Table~\ref{observation_log}.
Table \ref{main_table_sed} presents the parameters obtained from our SED fitting, such as parallax-based masses, luminosities, radii, and interstellar extinction. The final columns of this table also report the theoretical masses and evolutionary lifetimes inferred from our interpolation routine (see Sect. \ref{interpolation}). Finally, Table \ref{main_table_kin} provides the results of our kinematical analysis, listing the Galactic velocity components ($U, V, W$) and the corresponding probabilities of membership in the thin disk, thick disk, and halo populations.
In this paper, we continue to divide our stars into groups based on the visual inspection classification scheme described in Paper I with the addition of intermediate and extreme helium classes, in order to maintain consistency. 
Five objects (GALEX\,J19498-2806, GALEX\,J191509.0-290311, HD\,319179, CD-39\,14181, and EC\,21494-7018) have spectra of insufficient quality to be modelled reliably. For completeness, we nevertheless provide their atmospheric, kinematic, and stellar parameters, assigning a quality flag of ‘B’ in all three tables; reliable parameters are indicated by a flag of ‘A’. These five objects were excluded from the mass distribution analysis presented in Sect. \ref{mass_distribution}.

\begin{table*}[]
\caption{Excerpt from the table of atmospheric parameters.}
\label{main_table_atm}
\renewcommand{\arraystretch}{1.2}
\centering
\begin{tabular}{ccccccccc}
\toprule\toprule
Name & RA & Dec & Class & 	\teff & \logg & \logy & $v_\mathrm{rad}$ & Flag \\
& & & & [K] & [cm s$^{-2}$] &  & [\kms] & \\
\midrule
PG0314+146 & 49.4084 & 14.7732 & eHe-sdO & $46321^{+803}_{-803}$ & $5.80^{+0.08}_{-0.08}$ & $+1.00^{+0.11}_{-0.11}$ & $-16^{+3}_{-3}$ & A \\
2MASSJ02065617+1438585 & 31.7341 & 14.6495 & sdB & $30375^{+341}_{-371}$ & $5.80^{+0.07}_{-0.06}$ & $-2.38^{+0.06}_{-0.06}$ & $+4^{+8}_{-8}$ & A \\
PG0215+183 & 34.5658 & 18.5272 & sdB & $27666^{+320}_{-336}$ & $5.90^{+0.07}_{-0.07}$ & $-2.98^{+0.11}_{-0.14}$ & $+45^{+7}_{-7}$ & A \\
... &  & ... &  & ... &  & ... &  & ... \\
$[$L92b$]$MarkA & 310.9969 & -10.7949 & eHe-sdO & $36921^{+531}_{-594}$ & $6.02^{+0.09}_{-0.08}$ & $+1.14^{+0.06}_{-0.06}$ & $-12^{+6}_{-6}$ & A \\
\bottomrule
\end{tabular}
\tablefoot{Uncertainties on the atmospheric parameters (\teff, \logg, and helium abundance) are systematic as calculated in Sect. \ref{systematic_uncertainties}. \\
The full version is available at the CDS.
}
\end{table*}

\begin{table*}[]
\caption{Excerpt from the table of SED and HRD parameters.}
\label{main_table_sed}
\renewcommand{\arraystretch}{1.2}
\centering
\begin{tabular}{cccccccc}
\toprule\toprule
Name & $M_{\mathrm{SED}}$ & $L_{\mathrm{SED}}$ & $R_{\mathrm{SED}}$ & $E$(44$-$55) & $M_{\mathrm{HRD}}$ & $	\tau_{\mathrm{HRD}}$ & Flag \\
   &  [\msun] & [\lsun] & [\rsun] & [mag] & [\msun] & [Myr]  & \\
\midrule
PG0314+146 & $0.64^{+0.14}_{-0.11}$ & $114.97^{+11.08}_{-10.10}$ & $0.166^{+0.005}_{-0.005}$ & $0.197^{+0.003}_{-0.003}$ & $0.57^{+0.03}_{-0.06}$ & $100.7^{+47.5}_{-18.0}$ & A \\
2MASSJ02065617+1438585 & $0.42^{+0.07}_{-0.06}$ & $13.86^{+1.00}_{-0.94}$ & $0.135^{+0.004}_{-0.003}$ & $0.077^{+0.002}_{-0.002}$ & $0.43^{+0.03}_{-0.03}$ & $274.5^{+120.9}_{-66.3}$ & A \\
PG0215+183 & $0.59^{+0.10}_{-0.09}$ & $10.75^{+0.83}_{-0.77}$ & $0.143^{+0.004}_{-0.004}$ & $0.160^{+0.006}_{-0.004}$ & $0.43^{+0.02}_{-0.03}$ & $285.3^{+73.7}_{-43.7}$ & A \\
... &  & ... &  & ... &  & ... &  \\
$[$L92b$]$MarkA & $0.41^{+0.09}_{-0.07}$ & $17.73^{+1.31}_{-1.27}$ & $0.103^{+0.002}_{-0.002}$ & $0.037^{+0.003}_{-0.003}$ & $0.41^{+0.01}_{-0.01}$ & $341.2^{+20.8}_{-20.2}$ & A \\
\bottomrule
\end{tabular}
\tablefoot{The full version is available at the CDS.}
\end{table*}

\begin{table*}[]
\caption{Excerpt from the table of kinematic/membership parameters.}
\label{main_table_kin}
\renewcommand{\arraystretch}{1.2}
\centering
\begin{tabular}{ccccccccc}
\toprule\toprule
Name & $U$ & $V$ & $W$ & $p_{\mathrm{thin}}$ & $p_{\mathrm{thick}}$ & $p_{\mathrm{halo}}$ & Membership & Flag \\
&    [\kms]  &  [\kms]  & [\kms] & & & &  &\\
\midrule
PG0314+146 & $-50.2^{+2.9}_{-2.7}$ & $219.0^{+2.3}_{-2.1}$ & $-33.9^{+2.0}_{-2.3}$ & 0.88 & 0.12 & 0.0 & thin & A \\
2MASSJ02065617+1438585 & $-19.5^{+4.4}_{-3.6}$ & $218.8^{+3.0}_{-3.2}$ & $-16.8^{+4.5}_{-4.2}$ & 0.97 & 0.03 & 0.0 & thin & A \\
PG0215+183 & $76.7^{+4.9}_{-5.0}$ & $190.8^{+3.5}_{-3.9}$ & $-12.1^{+4.2}_{-4.7}$ & 0.69 & 0.31 & 0.0 & thin & A \\
... &  & ... &  & ... &  & ... &  & ... \\
$[$L92b$]$MarkA & $37.4^{+4.1}_{-4.2}$ & $261.7^{+3.7}_{-3.5}$ & $-19.8^{+2.9}_{-2.9}$ & 0.98 & 0.02 & 0.0 & thin & A \\
\bottomrule
\end{tabular}
\tablefoot{The full version is available at the CDS.}
\end{table*}

\subsection{Effective temperature and surface gravity}
\label{teff_logg}

The distribution of our 253 single-lined hot subdwarf stars in the effective temperature - surface gravity plane (hereafter Kiel diagram) is displayed in Fig. \ref{kiel} where the different spectral classes (see Sect. \ref{classification} and paper I) are colour-coded accordingly. 
The light-blue region encapsulates the EHB between the zero-age EHB (ZAEHB) at core helium ignition and the terminal-age EHB \citep[TAEHB;][]{Dorman_1993ApJ...419..596D_models} both for solar metallicity. The zero-age helium-burning main sequence \citep[He-MS;][]{Paczynski1971} is given as a dashed-grey line. 
Three post-EHB evolutionary tracks from \citet{Han_2002} are shown in black for the same core mass of 0.45\,\msun\ with envelope masses of 0.0, 0.001, and 0.005\,\msun, where the thickness of the line is proportional to the timescale. A double He-WD merger track from \citet{Zhang_Jeffery_2012MNRAS.419..452Z} for two $0.3$ \msun\ WDs is given as a orange dot-dashed line. TYC\,6800-72-1 appears to lie high on this track; however, it is unlikely to be physically associated with it, as it exhibits radial-velocity variations indicative of a close compact companion and is the only helium-enriched object in our sample with a detected companion (Dawson et al. in prep.).

In Fig.\ \ref{kiel}, most sdB and sdOB stars (blue stars and green diamonds, respectively) lie in a well-defined region between the ZAEHB and the TAEHB. Most sdBs appear to cluster where the EHB evolutionary tracks predict the longest timescale near the ZAEHB, giving rise to a dearth of sdBs near the TAEHB. 
The sdOBs predominantly reside at hotter temperatures (\teff\ ~$ > 30\, 000$~K) and higher surface gravities (\logg\ $> 5.5$). Nearly all of the eHe-sdOs, given by the filled orange circles, are located above the TAEHB and the He-MS, hinting at under-estimated surface gravities, though they align well with the He-WD merger track. 
Other large-sample studies of hot subdwarfs \citep[e.g.][]{Stroeer_2007A&A...462..269S, Latour_2014ApJ...795..106L, Fontaine_2014ASPC..481...83F, Nemeth_2012MNRAS.427.2180N, Luo_2019ApJ...881....7L, Luo_2021, Lei_2018ApJ...868...70L} have found substantial scatter in \logg\ around the He-MS.    
This scatter likely arises from the challenges inherent in modelling helium-enriched atmospheres, and highlights the need for improved atmospheric models next to homogeneous analysis techniques to resolve these differences. 
The hydrogen-rich sdOs (purple triangles) are primarily situated below the He-MS at higher temperatures. Their position aligns well with post-EHB tracks, reinforcing a connection to the sdB stars.

\subsection{Helium abundance}

Figure \ref{helium} shows that our sample follows the helium patterns reported in previous studies \citep{Edelmann_2003A&A...400..939E, Nemeth_2012MNRAS.427.2180N, Luo_2016, Lei_2018ApJ...868...70L}, with two branches of He-poor stars exhibiting increasing helium abundance toward higher \teff. The empirical relations from \citet{Nemeth_2012MNRAS.427.2180N} and \citet{Edelmann_2003A&A...400..939E} are included for reference as dashed black and red lines, respectively. The upper branch in our sample agrees reasonably well with the relation of \citet{Edelmann_2003A&A...400..939E}, but their lower branch is not clearly reproduced. In addition, most of our He-weak sdO stars lie below the relation reported by \citet{Nemeth_2012MNRAS.427.2180N}, but are of similar abundance to most of the sdB and sdOB stars.

Dashed-horizontal lines mark the \logy\ $=-1.2$ and \logy\ $=+0.6$ which we use to distinguish the iHe-sdOBs and the eHe-sdOB stars. A clear gap is seen at \logy\ $=-1.2$ but not for  \logy\ $=+0.6$.
Our sample contains one iHe-sdB \citep[CPD-20\,1123;][]{Vennes_2007ApJ...668L..59V}, which is marked as a black circle in the diagram; another peculiar star is the well-studied BD$+$75\,325 \citep{Gould_1957PASP...69..242G} as the only iHe-sdO in our sample. 

\cite{Nemeth_2012MNRAS.427.2180N} further discovered that hot subdwarfs predominantly cluster into two distinct groups within both the \teff\ - \logg\ and \teff\ - \logy\ diagrams. They inferred the presence of two characteristic hydrogen envelopes, each with differing masses and compositions. However, no clear gap is seen in our sample.

\subsection{Stellar parameters and the HRD}
\label{stellar_params}
The Kiel diagram provides useful information on the evolutionary stages and relative positions of stars in our sample. By combining this with precise parallax measurements from \textit{Gaia}, we can compute stellar luminosities and construct a Hertzsprung-Russell diagram (HRD), which allows for a more accurate characterisation of the stellar population thanks to precise luminosities. In Fig. \ref{hrd}, the HRD of our sample is shown, including the same evolutionary tracks, EHB, and He-MS features highlighted in Fig. \ref{kiel}.

Most sdB and sdOB stars are well situated in a narrow luminosity band between $\log L/L_{\odot} = 1.0$ and $1.5$, which implies that they are in a core helium burning stage. The horizontal dashed black lines indicate the solar metallicity zero-age EHB at $\log L/L_{\odot} = 1.05$ at the start of the $0.45$ \msun\ evolutionary tracks, and the point of core helium depletion for the same tracks. 
A few stars show much higher luminosities and are likely in a helium shell burning stage, such as the well-known post-HB star HD\,76431 \citep{Ramspeck_2001A&A...379..235R, Khalack_2014MNRAS.445.4086K}, as annotated in the plot, or are evolving towards the EHB.
All helium-enriched hot subdwarfs in our sample are remarkably close to the zero-age He-MS of \citet{Paczynski1971}, irrespective of whether they belong to the iHe or eHe class, suggesting that most are in a core-helium burning phase. 

Notably, excluding three objects with high re-normalised unit weight error values from \textit{Gaia} (RUWE; Sect. \ref{low_mass_sdbs}), as well as the pre-ELM candidates CPD-20\,1123 and EVR-CB-001 (see Sect. \ref{low_lum_sdbs}), 22 sdB stars are underluminous and are located below the canonical EHB in both the Kiel and HR diagrams (Figs. \ref{kiel}, \ref{hrd}), corresponding to a ratio of $0.10\pm 0.02$ (binomial uncertainty) relative to the number of stars on the EHB band (sdB$+$sdOB). 
These stars are marked with black pentagons in all three plots (Fig. \ref{kiel_helium_hrd}) for comparison. 
Interestingly, around half of these objects have exceptionally low helium abundances \citep[\logy\ $\leq -3.8$; also seen by][]{latour2025arizonamontrealspectroscopicsurveyhot}, similar to the sdB closest to the Sun, HD\,188112, which is also annotated in Fig. \ref{kiel_helium_hrd}. 
These stars are classified as sdBs because their spectra are void of ionised helium lines upon visual inspection. Most have cooler temperatures and lower helium abundances than those identified in this region previously by \citet{Edelmann_2003A&A...400..939E}, whose trends marked with dashed-red lines do not extend to \logy\ $\leq -3.8$.

\subsubsection{Underluminous hot subdwarfs}
\label{low_lum_sdbs}
The physical origin of the underluminous population in our sample remains uncertain. These stars are located below the EHB and exhibit systematically higher surface gravities at a given effective temperature compared with the bulk of the sdB population. 
The horizontal red-dashed lines in Fig. \ref{hrd} mark the ZAEHB for the $0.40$ \msun\ and $0.35$ \msun\ evolutionary tracks from \cite{Han_2002}. We find that these stars are likely associated with these low-mass tracks and discuss this further in Sect. \ref{interpolation}.

We define `below-EHB' objects as those with $\log (L/L_{\odot}) \leq 1.05$, based on the solar-metallicity $0.45$ \msun\ EHB tracks of \citet{Han_2002}, which is appropriate given the local nature of our sample. The below-EHB population have been identified previously \citep[e.g.][]{Geier_2022, He_2025A&A...693A.121H}, though differ subtly in the criterion used to define them.  
At $\log L/L_{\odot} \leq 1.05$, we also observe a sharp drop in the luminosity distribution of the sdB and sdOB stars (Fig. \ref{kde_mass_luminosity}), marking the possible transition to the low-mass hot subdwarfs, formed by non-degenerate helium ignition \citep[e.g.][]{Arancibia_Rojas_2024MNRAS.52711184A}. More discussion on the link between the low-luminosity stars and their low-mass is presented in Sect. \ref{intermediate_mass_progenitors}.

Other observational studies in the literature generally define the EHB using metal-poor $0.47$ \msun\ evolutionary tracks, such as those of \citet{Han_2002} or \citet{Dorman_1993ApJ...419..596D_models}. These tracks lie at higher luminosities in the HRD compared to the ZAEHB shown in Fig.~\ref{hrd}. This difference arises because metal-poor stars have lower opacities, allowing their cores to cool more efficiently. As a result, a larger amount of helium must accumulate before ignition, leading to a more massive core, a condition that also holds for more massive progenitors \citep[$M_{\mathrm{ZAMS}} \geq 2.0$\,\msun ;][]{Arancibia_Rojas_2024MNRAS.52711184A}. 
If metal-poor tracks were adopted here, the proportion of below-EHB objects in our sample would thus be even higher. 
Another consideration is that the masses at the ZAEHB depend on the physics assumed in the models, such as overshooting prescriptions \citep[see e.g.][for more details]{Ostrowski_2021MNRAS.503.4646O, Xiong_2017A&A...599A..54X}.

Among the underluminous hot subdwarfs is HD\,188112, regarded as the prototype sdB extremely low-mass (ELM) white dwarf progenitor \citep{Heber_2003A&A...411L.477H} and the closest object in our sample to the Sun at just 71~pc. As already pointed out in Sect. \ref{systematic_uncertainties}, metallicity has an impact on the derived atmospheric and stellar parameters. 
Because \cite{Latour_2016A&A...585A.115L} analysed UV data, HD\,188122 has a well-determined abundance pattern, which turns out to be metal-poor. 
This pattern is well represented by $\log z/z_\mathrm{sdB}=-1$, which we adopted to derive \teff\ $=21650\pm350$ K and \logg\ $=5.81\pm0.07$ for HD\,188112 from a very high SNR WHT/ISIS spectrum displayed in Fig. \ref{sdB_spectrum}. 
In particular \logg\ is higher than derived in the study of \citet{Heber_2003A&A...411L.477H} which gave \teff\ $=\,21500\pm500$ K and \logg\ $=\,5.66\pm0.05$. The difference in \logg\ can be traced back to a combination of different model atmospheres used at the time \citep[see][]{Heber_2000A&A...363..198H}, and the wavelength coverage of the spectra analysed. Notably, the TWIN spectra analysed in \citet{Heber_2003A&A...411L.477H} do not adequately cover the hydrogen Balmer jump, which is crucial for accurate \logg\ determination. 
The updated parameters are shown in the Kiel (Fig. \ref{kiel}) and HR (Fig. \ref{hrd}) diagrams, with the revised position of HD\,188112 indicated by a light dashed-grey line, and its previous location is marked with a grey star. Because its effective temperature is essentially unchanged, its HRD position remains the same. In this diagram, the star lies very close to the low-mass sdB ($0.35\,$\msun) tracks of \citet{Han_2002}. Given the higher \logg\ found in this work, the mass derived from the SED and parallax is $0.33\,$\msun, substantially higher than the previously inferred $0.24$ \msun\ \citep{Heber_2003A&A...411L.477H}, but in excellent agreement with the evolutionary tracks. 
Hence, we find it very likely that HD\,188112 is not an ELM progenitor as previously thought, but among the least massive EHB stars identified in this sample. 
TYC\,5417-2552-1 (annotated in Fig. \ref{kiel_helium_hrd}) has very similar atmospheric and stellar properties to HD\,188112. We find TYC\,5417-2552-1 has a mass of $0.33^{+0.03}_{-0.03}$ which aligns well with the low-mass tracks of \citet{Han_2003} and is therefore also likely a low-mass sdB rather than an ELM progenitor as proposed by \citet{Kawa_2015}. 

CPD-20\,123 is the only identified iHe-sdB star within the 500 pc sample and is marked with a black circle in Fig. \ref{kiel_helium_hrd}. This object has previously been interpreted as a post-common-envelope binary evolving towards the EHB, likely hosting either a WD or dM companion on a 2.3 day orbital period \citep{Naslim_2012MNRAS.423.3031N, Kupfer_2015}. However, our analysis yields a substantially lower mass for the hot subdwarf than assumed in these studies, $0.18^{+0.02}_{-0.02}$ \msun, based on its atmospheric parameters (\teff\ $=\,21610\pm350$ K and \logg\ $=\,5.04\pm0.04$) and the \textit{Gaia} parallax. 
This suggests that CPD-20\,1123 might instead be a pre-ELM object evolving towards the white dwarf cooling track, rather than an EHB progenitor. 
The system may thus be similar to the pre-ELM + He-WD system EVR-CB-001 \citep{Ratzloff_2019ApJ...883...51R}, which is also part of the 500~pc sample; in this case our mass ($0.30^{+0.06}_{-0.05}$ \msun) is somewhat higher than the previously published value ($0.20^{+0.05}_{-0.05}$ \msun), but still consistent with a pre-ELM nature.
Both TYC\,5417-2552-1 and CPD-20\,123 are removed from subsequent analysis. 

\subsubsection{Hot and helium-poor sdBs}
Three stars with helium abundances below \logy\ $<-3.8$ (UCAC4\,314-235772, PG\,1538+401, and TYC\,4651-1475-1) lie between 35\,000 and 40\,000 K, a temperature region of extreme helium abundance diversity in hot subdwarf stars. These objects show no evidence of ionised or neutral helium in their atmospheres, and are therefore also classified as sdBs despite having effective temperatures close to the coolest sdO stars. 
In Fig. \ref{kiel_helium_hrd}, they are distinguished with crimson pentagons and unlike the cool sdBs with \logy\ $<-3.8$ discussed above, are located above the ZAEHB in both Fig. \ref{kiel} and \ref{hrd}, consistent with an evolved state. This may hint at a possible evolutionary connection between the cool helium-poor sdBs and the helium-poor sdOs.

\begin{figure}
    \centering
    \includegraphics[width=1\linewidth]{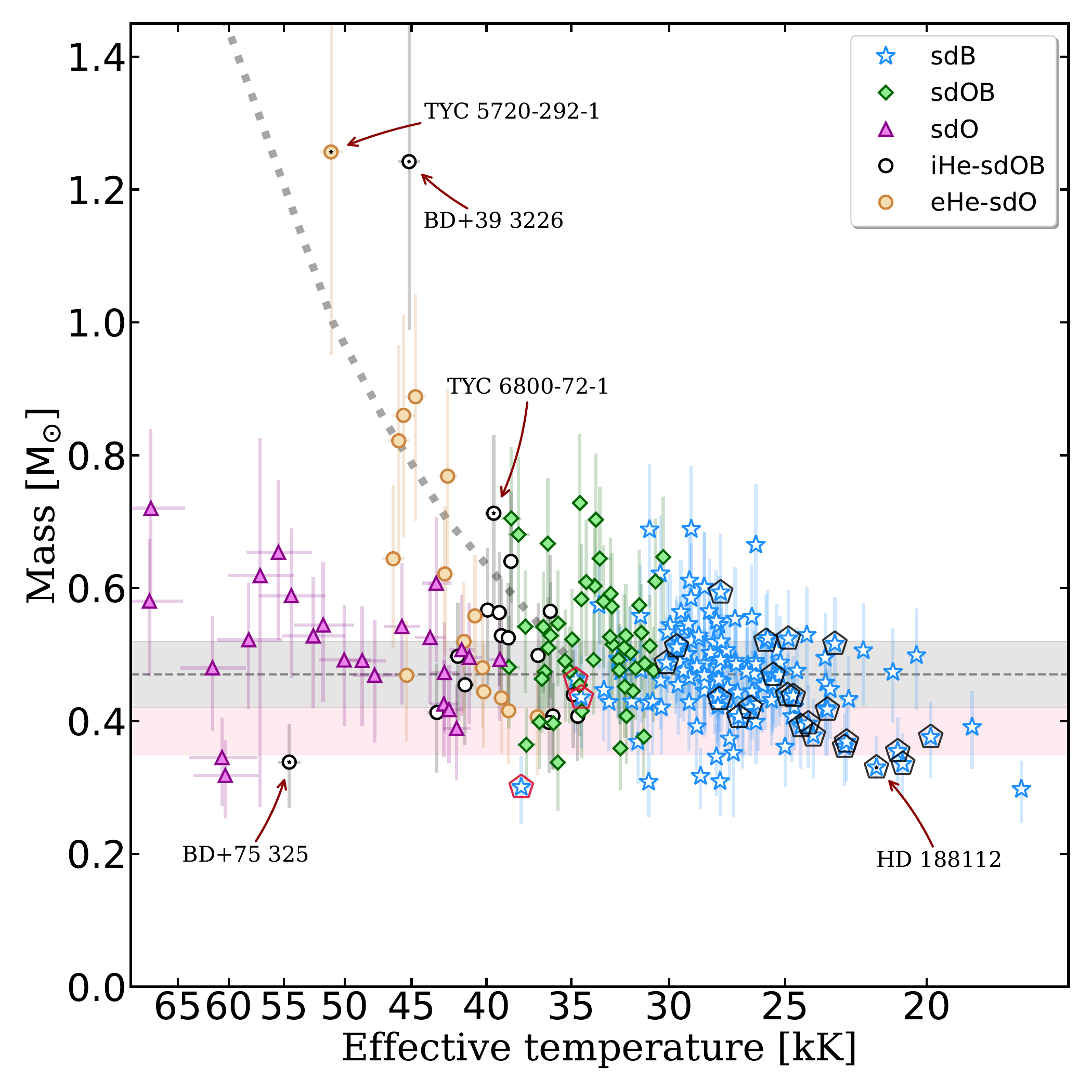}
    \caption{Distribution of our stars in the mass - effective temperature plane. The He-MS \citep{Paczynski1971} is shown as a dashed grey line. The grey-shaded band is centered on a mass of $0.47$ \msun\ and spans $\pm0.05$ \msun, corresponding to our mean uncertainties. The pink-shaded region extends down to $0.35$ \msun, the minimum core mass predicted by the evolutionary tracks of \citet{Han_2002}. Pentagons indicate the same stars as in Fig.~\ref{kiel_helium_hrd}.}
    \label{mass_temperature}
\end{figure}

\begin{table}[]
\centering
\caption{Median masses and luminosities from the SED/parallax and HRD.}
\label{median_masses}
\renewcommand{\arraystretch}{1.3}
        \begin{tabular}{cccc}
                \toprule
                \toprule
                Class &  $M/M_{\odot}$ & $M/M_{\odot}$ &  $\log L/L_\odot$      \\
                 & (SED) & (HRD) & (SED)  \\
                \midrule
                 All stars & $0.48^ {+0.14}_{-0.10}$ & $0.47^ {+0.08}_{-0.05}$  &   $1.31^ {+0.40}_{-0.20}$ \\
                 sdB & $0.47^ {+0.11}_{-0.09}$ & $0.45^ {+0.04}_{-0.05}$ &    $1.21^ {+0.17}_{-0.16}$   \\
                 sdOB & $0.51^ {+0.14}_{-0.11}$ & $0.48^ {+0.08}_{-0.05}$ &   $1.35^ {+0.15}_{-0.14}$  \\
                 sdO & $0.50^ {+0.15}_{-0.12}$ & $0.56^ {+0.12}_{-0.10}$ &   $1.92^ {+0.22}_{-0.34}$   \\
                 iHe-sdOB & $0.50^ {+0.18}_{-0.12}$  & $0.53^ {+0.14}_{-0.09}$ &   $1.54^ {+0.47}_{-0.26}$ \\
                 eHe-sdO & $0.58^ {+0.33}_{-0.16}$ & $0.52^ {+0.08}_{-0.07}$  &    $1.88^ {+0.24}_{-0.23}$  \\
                 iHe-sdOB (He-MS)$^\dag$ & -- & $0.59^ {+0.08}_{-0.07}$ &   --  \\
                 eHe-sdO (He-MS)$^\dag$ & -- & $0.70^ {+0.12}_{-0.07}$  &    -- \\                
                \midrule
                Below EHB & $0.43^ {+0.11}_{-0.09}$  & $0.39^ {+0.03}_{-0.05}$ & $0.94^ {+0.09}_{-0.21}$ \\
        \bottomrule
        \end{tabular}
        \tablefoot{$^\dag$\citet{Paczynski1971}: theoretical masses derived by projecting these stars to their nearest positions on the He-MS in the HRD.\\
        Uncertainties correspond to the 16th-84th percentiles, as described in Sect.~\ref{mass_distribution}.
        }

\end{table}

\begin{figure*}
    \centering
    \begin{subfigure}[b]{0.32\linewidth}
        \centering
        \includegraphics[width=\linewidth]{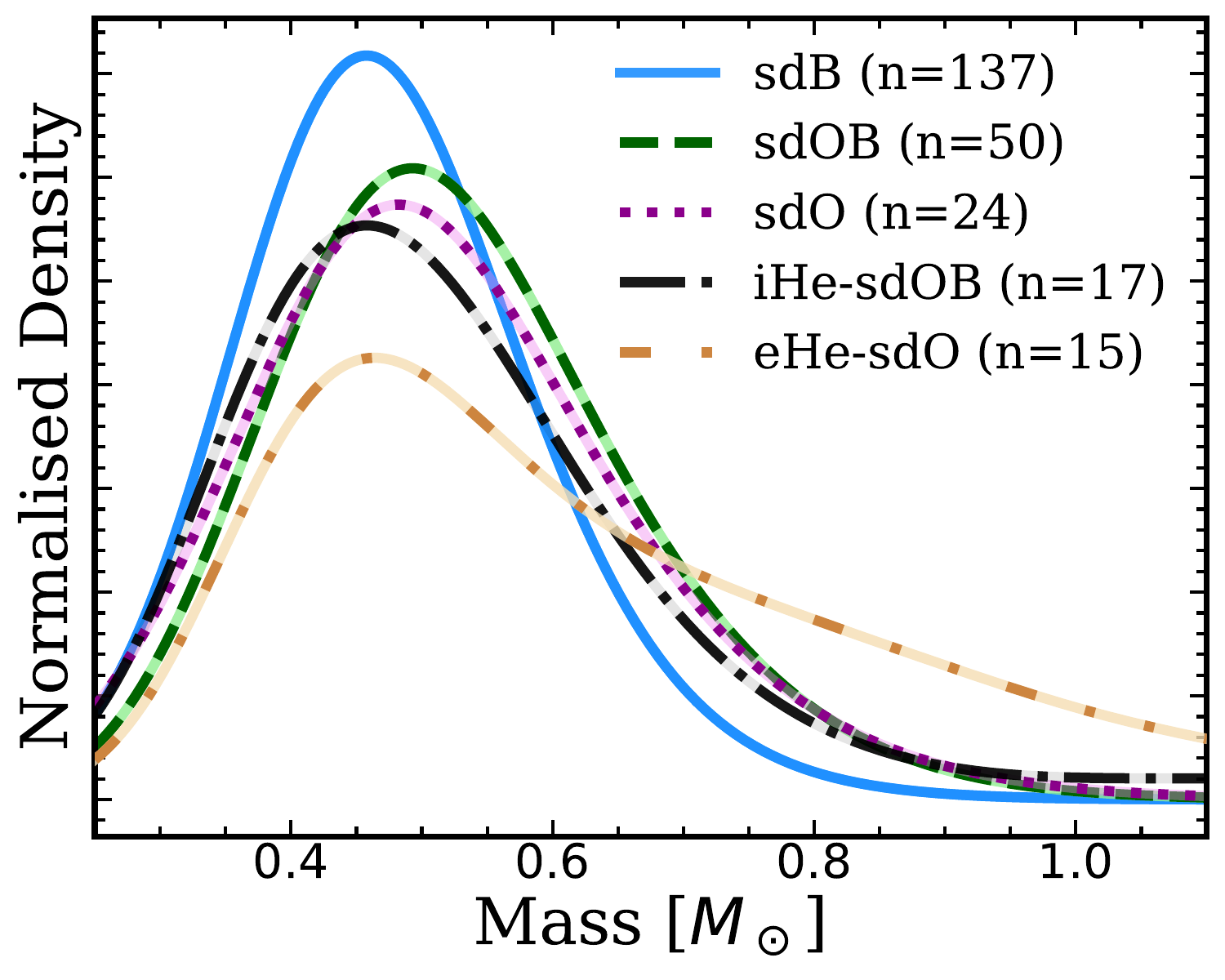}
    \end{subfigure}
    \hspace{0.01\linewidth}
    \begin{subfigure}[b]{0.32\linewidth}
        \centering
        \includegraphics[width=\linewidth]{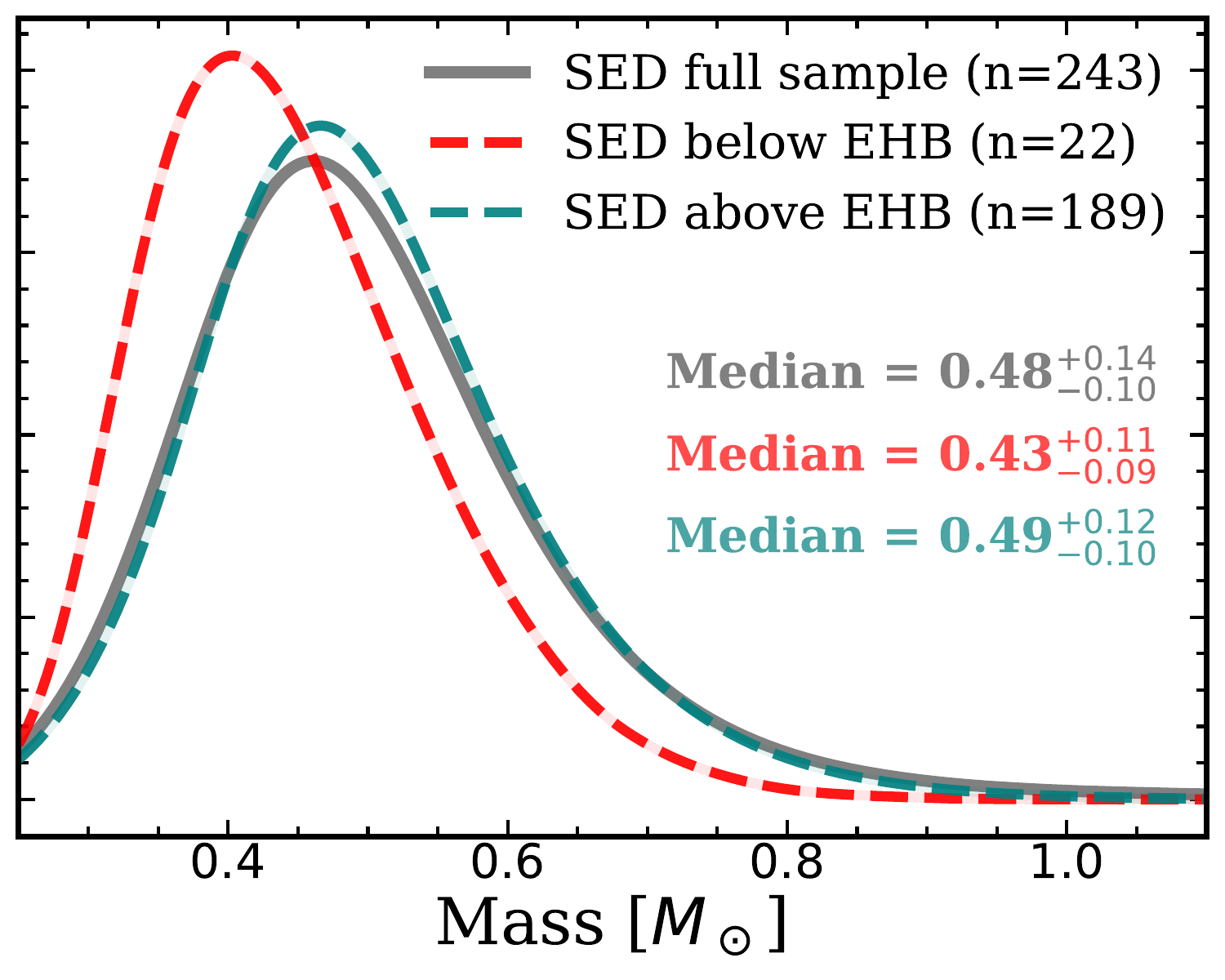}
    \end{subfigure}
    \hspace{0.01\linewidth}
    \begin{subfigure}[b]{0.32\linewidth}
        \centering
        \includegraphics[width=\linewidth]{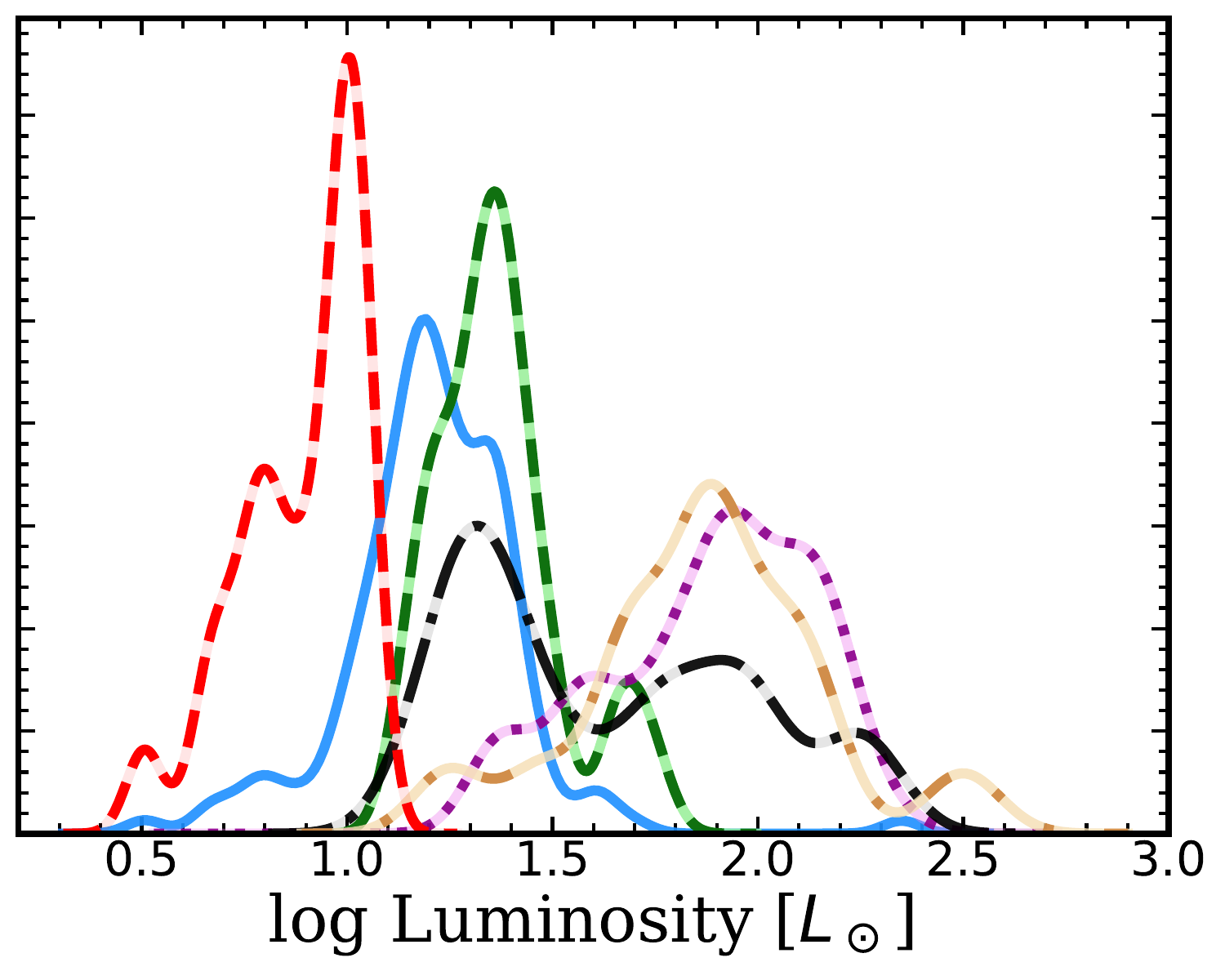}
    \end{subfigure}
    \caption{KDE representation of the mass and luminosity distributions for the full sample and subgroups of hot subdwarf stars. Uncertainties are not included in the KDEs, allowing clearer comparison between different hot subdwarf classes.}
    \label{kde_mass_luminosity}
    \vspace{5pt}
    \centering
    \begin{subfigure}[b]{0.32\linewidth}
        \centering
        \includegraphics[width=\linewidth]{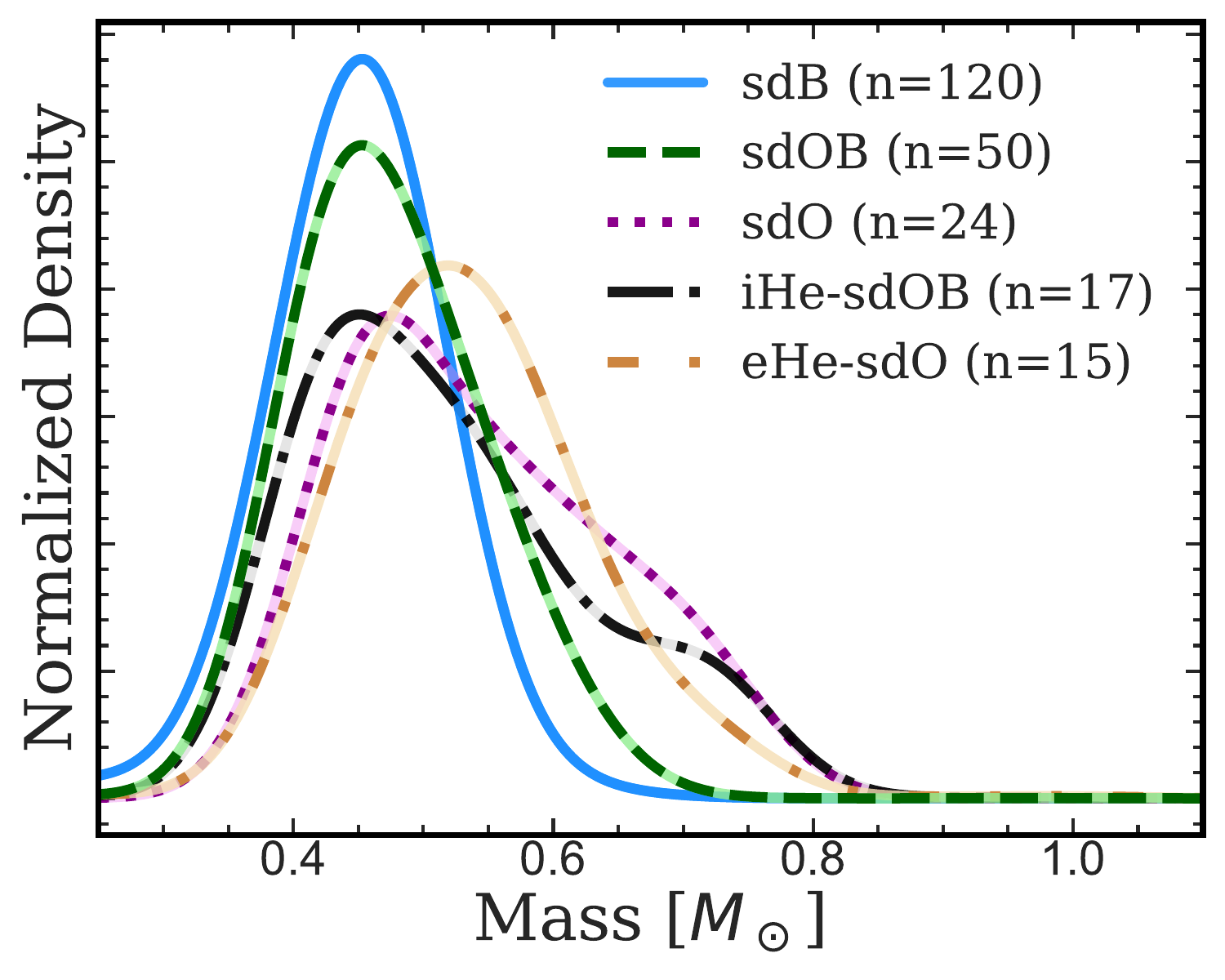}
    \end{subfigure}
    \hspace{0.01\linewidth}
    \begin{subfigure}[b]{0.32\linewidth}
        \centering
        \includegraphics[width=\linewidth]{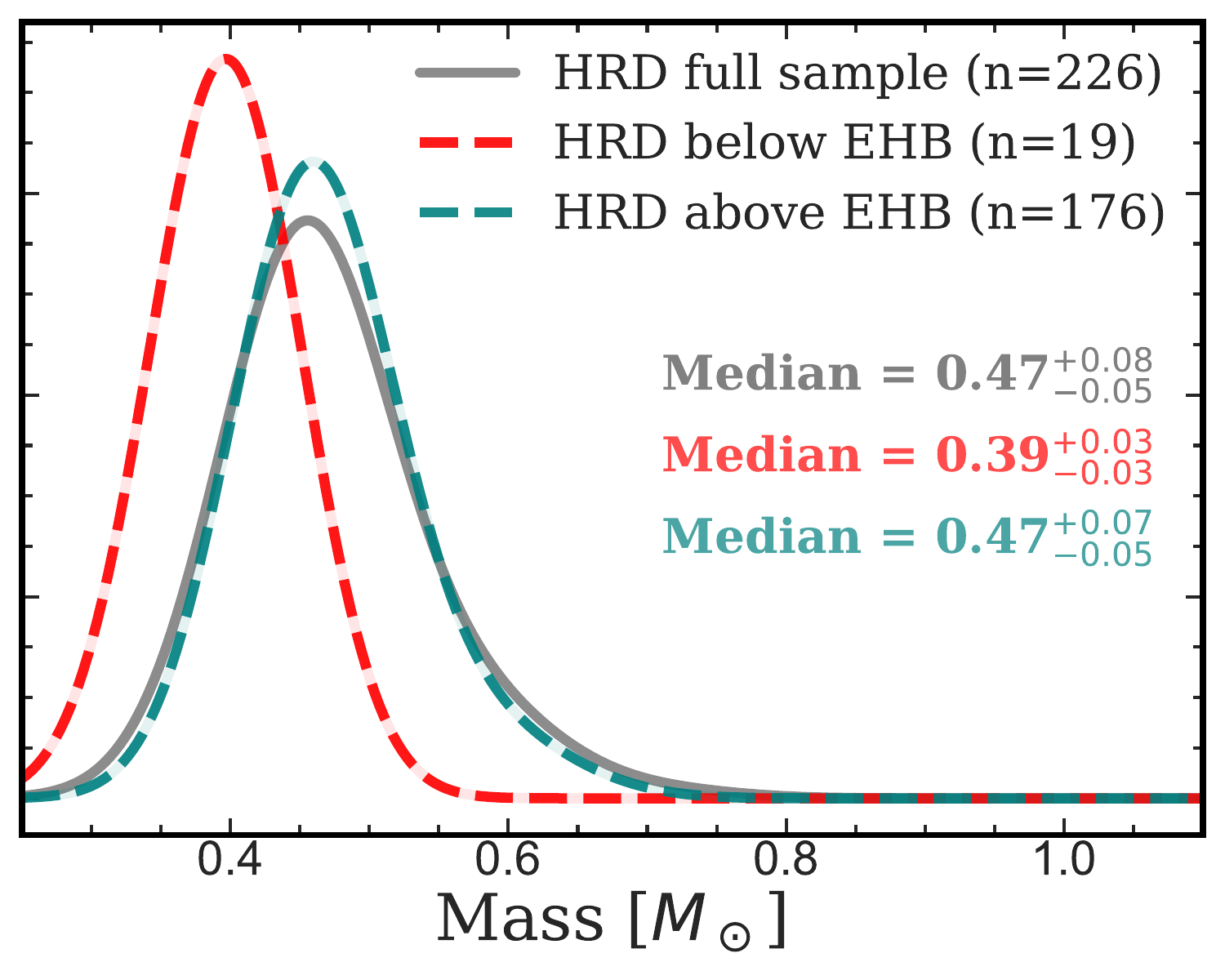}
    \end{subfigure}
    \hspace{0.01\linewidth}
    \begin{subfigure}[b]{0.32\linewidth}
        \centering
        \includegraphics[width=\linewidth]{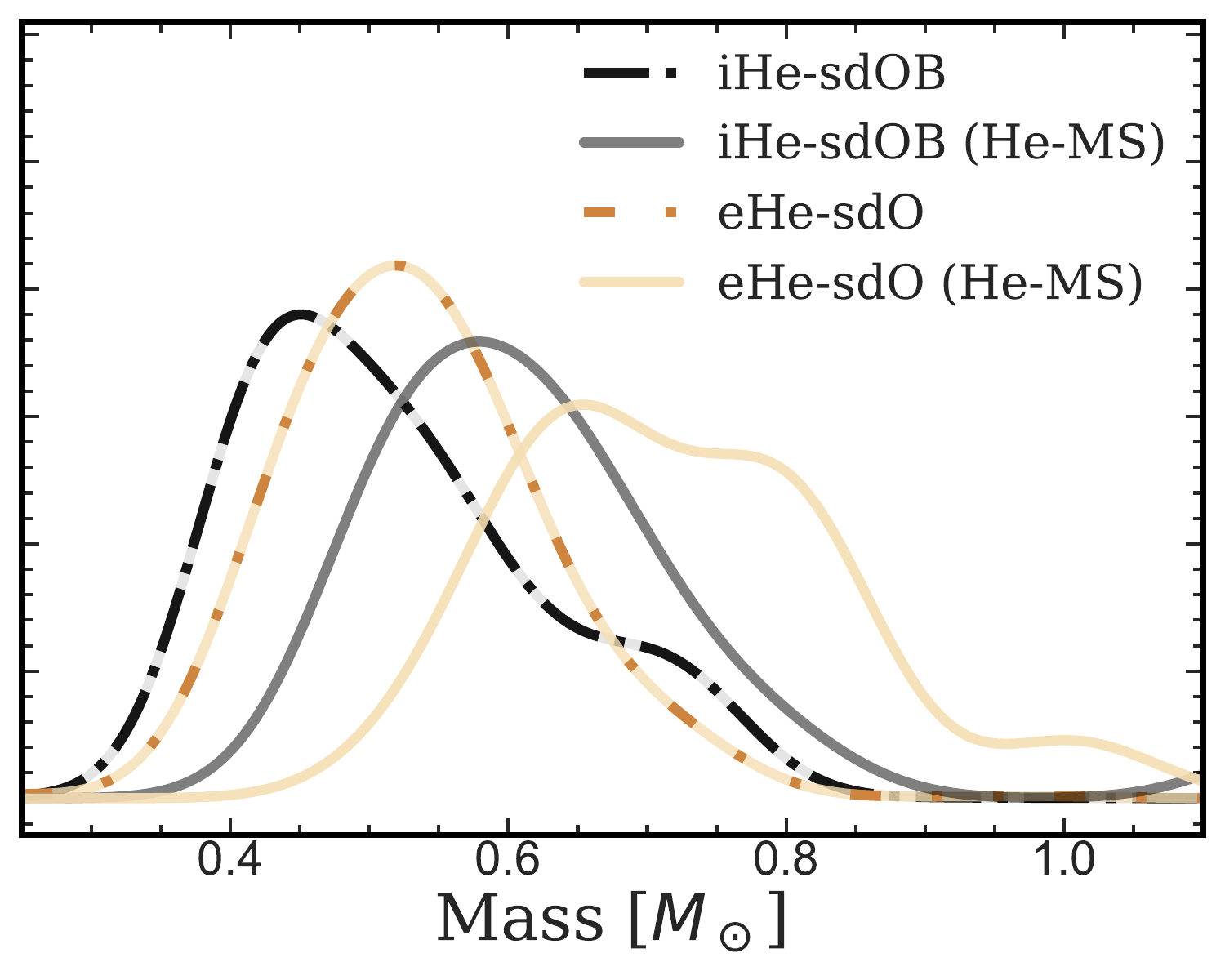}
    \end{subfigure}  
    \caption{Same KDEs as presented in Fig. \ref{kde_mass_luminosity}, but using theoretical masses derived from our MCMC interpolation procedure in the HRD described in Sect. \ref{interpolation}. Additionally, the right panel shows the estimated mass distributions of the He-rich stars derived by projecting each object to its nearest point along the He-MS \citep{Paczynski1971} in the HRD.}
    \label{kde_mass_luminosity_mcmc}
\end{figure*}

\subsection{Mass and luminosity distributions}
\label{mass_distribution}

Our results from the SED fitting routine detailed in Sect. \ref{sed_fitting} combined with \textit{Gaia} DR3 parallaxes are used to calculate the median masses and luminosities for each spectral class and are listed in Table \ref{median_masses}. The distribution of masses as a function of \teff\ is shown in Fig.\ \ref{mass_temperature}, where the black shaded area highlights $\pm\, 0.05\,$\msun\ around a mass of $0.47\,M_{\odot}$. The pink-shaded region extends this further down to $0.35\,M_{\odot}$, which is the minimum core mass covered by the evolutionary tracks of \citet{Han_2002}. Our results are further presented as mass and luminosity distributions in Fig. \ref{kde_mass_luminosity}, shown as normalised Kernel Density Estimates (KDEs) for a clear comparison between sub-classes. A KDE is a smoothed representation of the distribution, where each data point contributes a small Gaussian kernel; the sum of these kernels gives a continuous estimate of the probability density. The smoothing bandwidth is $0.05$\,\msun\ in all cases which sets the scale over which individual measurements contribute to the distribution. The uncertainties are derived directly from simple Monte Carlo arrays, with asymmetry reflecting the 16th and 84th percentiles.
For the full sample (flag 'A' in table \ref{main_table_sed}) of 243 single-lined stars (grey line; central panel of Fig.\ \ref{kde_mass_luminosity}), we derive a median mass of $M =\,0.48^ {+0.14}_{-0.01}$\msun, which is within the range of previous studies \citep[$\sim\,0.46 -0.5\,M_\odot$][]{Fontaine_2012,Schaffenroth_2022_1}, depending upon metallicity and evolutionary model prescriptions \citep[see][and references therein]{Sweigart_Gross_1978ApJS...36..405S, Cassisi_2016MmSAI..87..332C, Arancibia_Rojas_2024MNRAS.52711184A}. 

\citet{Lei_2023ApJ...953..122L_mass} analysed the mass distributions of 667 single-lined hot subdwarf stars observed with LAMOST, dividing them into three groups based on parallax uncertainty (see their table 2). For comparison with the 500~pc sample, their group 3 is most relevant, both in terms of sample size and parallax precision (better than 5\,\%). Their overall mean mass of $0.48\,M_\odot$ agrees with our median result of $0.48\,M_\odot$. Similar agreement is found for He-rich hot subdwarfs, which show a relatively lower peak mean mass than expected \citep{Han_2003} of $0.5\,M_\odot$ in both studies, highlighting the need for improved modelling of the ionised helium-dominated atmospheres. 
However, a notable discrepancy exists for the hydrogen-rich sdO stars: \citet{Lei_2023ApJ...953..122L_mass} report a mean of $0.36\,M_\odot$, significantly lower than our $0.50\,M_\odot$, which is unexpected given the likely evolutionary connection between sdB and sdO stars. 

\subsection{Low-mass hot subdwarfs}
\label{low_mass_sdbs}
The central panel of Fig. \ref{kde_mass_luminosity} shows the mass distribution of underluminous sdB stars located below the EHB in the HRD (22 stars; $\log L/L_{\odot}$ $<$ 1.05) in red compared to the sdB, sdOB, and sdO stars above the EHB (189 stars; $\log L/L_{\odot}$ $>$ 1.05) in teal. The below-EHB objects show an offset, peaking at $M\,=\,0.43^ {+0.11}_{-0.09}\,M_{\odot}$. 
\citet{Lei_2023ApJ...953..122L_mass} noted that large parallax uncertainties (up to about $20\%$) can bias samples toward low-mass objects, likely due to systematic \textit{Gaia} effects at greater distances. Our sample is largely unaffected, with a mean parallax uncertainty of only $1.54\%$. The exceptions are EC\,20106-5248, PG\,1634+061, and RL\,105, with parallax uncertainties of 3 to 6\,\% and high RUWE values ($2$ to $3$). Their parallax-derived masses fall below $0.4\,M_{\odot}$ and are likely unreliable due to poor \textit{Gaia} astrometry, while their spectroscopic surface gravities place them on the EHB and should not be affected. Their stellar parameters are provided in tables \ref{main_table_sed} and \ref{main_table_kin} with the flag 'B', but are excluded from the mass distribution and HRD analysis in this paper. 

To evaluate the statistical significance of the low-mass peak observed for our underluminous hot subdwarfs, we apply a two-sample Kolmogorov-Smirnov (KS) test, supplemented by the Mann-Whitney $U$ test as a complementary check, both implemented through the \texttt{SciPy} Python module. The underluminous population are compared against those classed as sdB, sdOB, and sdO with $\log L/L_{\odot}$ $>$ 1.05 (helium-rich objects are excluded). 
We accounted for measurement uncertainties in the parallax-derived masses using Monte Carlo resampling. In 10\,000 trials, each star’s mass was randomly varied within its uncertainty, and the statistical tests were repeated. Both the KS and Mann–Whitney tests consistently yielded low $p$-values (median $p_{\mathrm{KS}}\sim0.02$, $p_{\mathrm{MW}}\sim0.007$), rejecting the possibility that the two samples share the same parent distribution with $\sim$99.5\% confidence. This provides strong evidence that the underluminous stars have a different mass distribution from the general hot subdwarf population in our sample.

\begin{table}

\centering
\caption{Kinematic population membership. }
\label{population_membership}
\renewcommand{\arraystretch}{1.13}
        \begin{tabular}{cccc}
                \toprule\toprule
                 & Thin & Thick & Halo  \\
                \midrule 
                 All stars & 224 & 22 & 4  \\
                  $[\%]$ & $86\pm2$ & $13\pm1$ & $1\pm1$  \\
                   \midrule          
                 sdB & 127 & 10 &   1 \\
                 sdOB & 47 & 5 & 0  \\
                 sdO & 23 & 1 & 0 \\
                 iHe-sdOB & 11 & 2 & 3 \\
                 eHe-sdO & 12 & 3 &  0 \\
        \bottomrule
        \end{tabular}
\end{table}

\begin{figure*}
    \centering
    \includegraphics[width=0.90\linewidth]{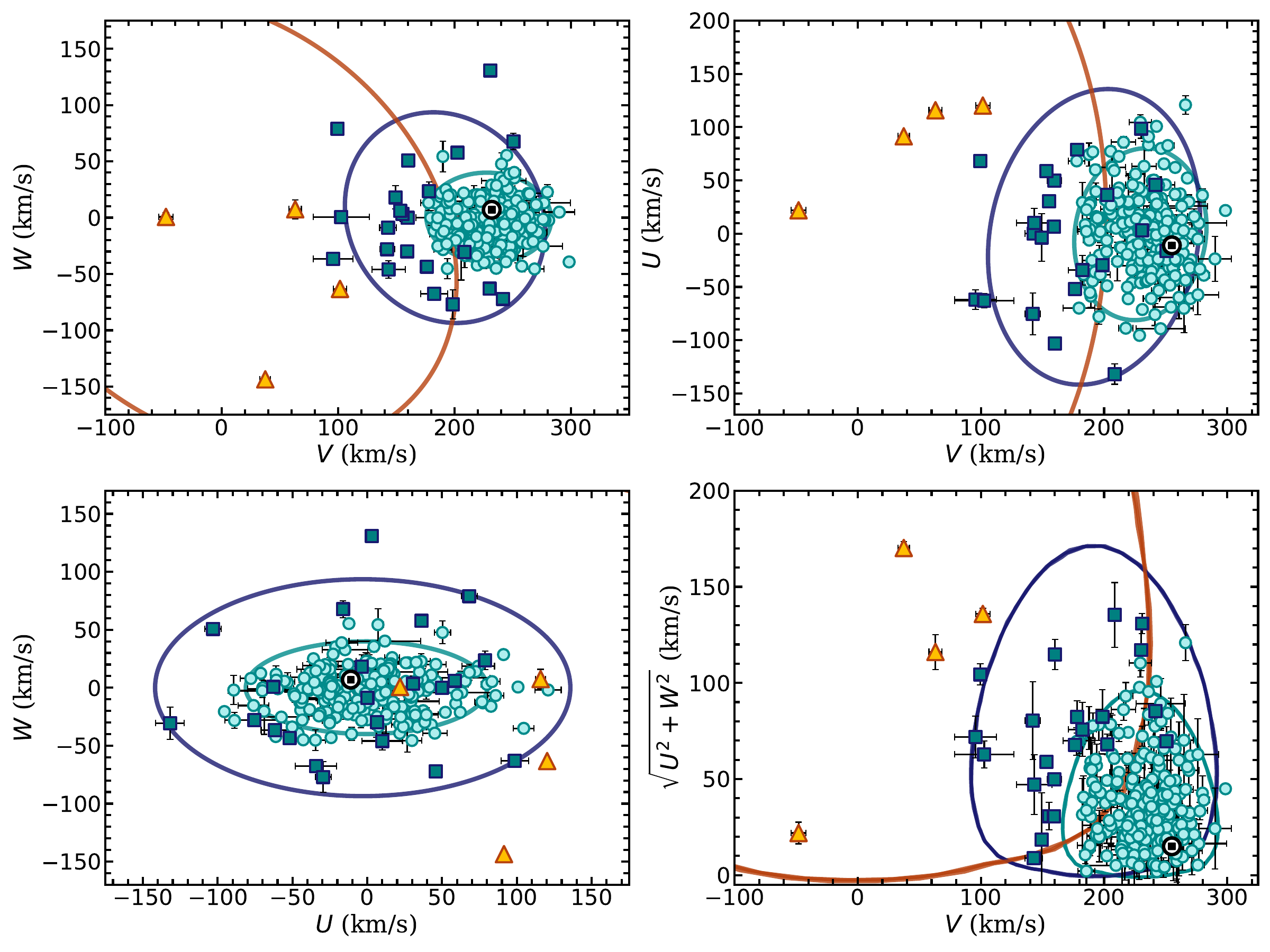}
    \caption{Distribution of the 500~pc sample in the $W$-$V$, $U$-$V$, $W$-$U$, and Toomre (lower right) diagrams. The stars are marked as light blue circles, dark blue squares, and orange triangles to indicate thin disk, thick disk and halo membership, respectively. Two-sigma contours of the Galactic components are given in the same colour scheme. The position of the Sun is given by the black circle. }
    \label{toomre_UVW}
\vspace{5pt}
    \centering
    \includegraphics[width=0.90\linewidth]{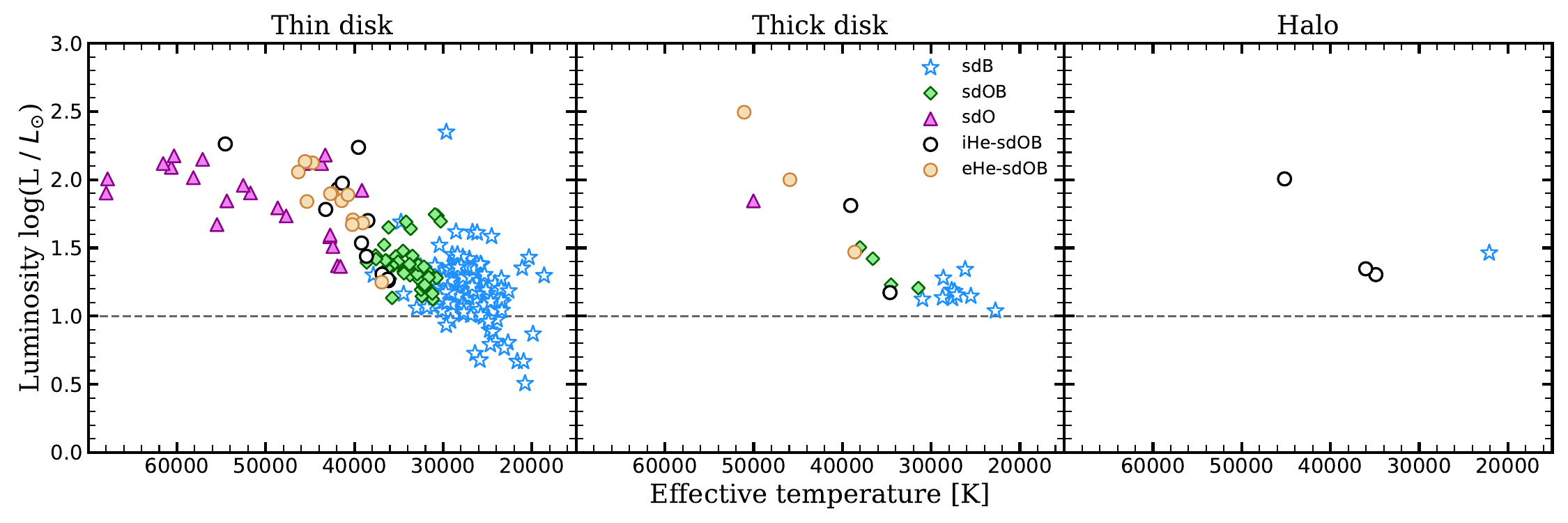}
    \caption{Three-panel plot showing the HRDs for the thin disk, thick disk, and halo populations in our sample, as determined from our kinematic analysis in Sect.~\ref{kinematics}. The Solar-metallicity ZAEHB at $L/L_{\odot} = 1.05$  is overlaid to highlight that only objects associated with the thin disk population fall below the EHB.}
    \label{hrd_thin_thick_halo}
\end{figure*}

\subsection{Population membership results}

The results of our kinematical analysis (described in Sect.\ \ref{method:kinematics}) are summarised in Table \ref{population_membership}, and visualised in the $U$-, $V$-, and $W$-component diagrams, along with the Toomre diagram in Fig.~\ref{toomre_UVW}.  
The pale-blue, dark-blue, and orange contours mark the 2$\sigma$ contours of the thin disk, thick disk, and halo populations, respectively. For visual purposes, individual stars are coloured based on the Galactic component membership simply with the highest probability. In table \ref{main_table_kin}, these numbers therefore add up to more than the quoted membership fraction, which are calculated by summing up each membership probability given in table \ref{main_table_kin}. As expected for a local neighbourhood sample, the majority ($86\pm 2\%$) of stars are classified as thin disk objects, while $13\pm 1 \%$ and $1 \pm 1 \%$ are associated with the thick disk and halo, respectively. 
Notably, three of the four stars assigned to the halo population are helium-enriched, namely LS\,IV\,-14\,116 \citep{Randall_2015A&A...576A..65R}, Feige\,46 \citep{Latour_2019A&A...623L..12L}, and BD+39\,3226 \citep{Schindewolf_2018} and are previously known to belong to the halo. If eHe-sdOs are indeed post-merger objects, they are expected to be on average older owing to their formation delay time \citep[see figure 14 in][]{Nicolas_2_2025PASA...42...12R}, and should therefore preferentially be in older Galactic populations and not in the young disk \citep{Han_2008A&A...484L..31H, Nicolas_2_2025PASA...42...12R}. 
The only helium-poor star in the halo, BD\,+48\,2721, is a peculiar object with an identified $^{3}\mathrm{He}$ isotopic anomaly, and is likely a BHB star given its relatively low effective temperature \citep{Schneider_2018}. 

Figure \ref{hrd_thin_thick_halo} presents the HRD for stars belonging to the thin disk, thick disk, and halo populations, arranged from left to right. The dashed-black line represents the solar ZAEHB at $\log L/L_{\odot} = 1.05$ which we have used to define the underluminous hot subdwarfs (Sect. \ref{low_lum_sdbs}). A striking feature is the exclusive presence of underluminous hot subdwarfs within the Galactic thin disk population. While the ZAEHB is expected to shift to slightly higher luminosities in older, more metal-poor populations such as the thick disk and halo \citep{Dorman_1993ApJ...419..596D_models}, this trend alone does not account for the absence of such low-luminosity stars in those populations. 
We note that the thick disk and halo subsamples comprise only 22 and 4 stars, respectively, limiting the statistical power of this comparison. Under the null hypothesis that all populations share a common below-EHB fraction, the probability of observing zero below-EHB stars in the thick disk is $p=0.15$ (binomial statistics), consistent with a $\sim$$1\sigma$ statistical fluctuation, and the halo result is not statistically significant. 
However, similar studies of hot subdwarf populations with comparable sample sizes found smaller below-EHB fractions of $6\%$ \citep{latour2025arizonamontrealspectroscopicsurveyhot} and $0\%$ \citep{Heber2025}, compared with $10\%$ in our sample. Although these surveys are not volume-limited, their targets have larger average distances of 800~pc and 1500~pc, respectively, supporting the conclusion that the presence of below-EHB hot subdwarfs is linked to the age of the stellar population.

\section{Discussion}
\label{discussion}
\subsection{Comparison to theoretical tracks}
\label{interpolation}

Accurate luminosities are available for all stars in our sample (see Sect.\ \ref{sed_fitting}). This enables us to interpolate theoretical stellar masses from their positions in the HRD, which we can then directly compare with the masses obtained from the SED and parallax fitting. 
For simplicity, we relied on the  evolutionary tracks from \citet{Han_2002}, which cover the EHB and early post-EHB evolution for core masses $M_\mathrm{core}$ between 0.35 and 0.75\,\msun\ and hydrogen-rich envelope masses $M_\mathrm{envelope}$ between 0 and 0.005\,\msun. 
Because the evolutionary tracks overlap considerably in the HRD, we used a Markov Chain Monte Carlo (MCMC) approach to explore the full range of possible age, core mass, and envelope mass combinations, and to predict $\log T_{\mathrm{eff}}$ and $\log L$ for each case. 
These predictions are compared to the spectroscopically derived effective temperature and parallax-luminosity within a multivariate normal likelihood function incorporating observational uncertainties and their covariance. \footnote{The covariance matrix between $\log T_{\mathrm{eff}}$ and $\log L$ was estimated from a sample of 50{,}000 Monte Carlo draws from the SED fitting posteriors. 
This accounts for observational uncertainties and correlations between parameters during the MCMC sampling.} 
The log-likelihood function is given by
\begin{equation}
\ln \mathcal{L} = -\frac{1}{2} \left( \chi_{\log T_{\mathrm{eff}}}^2 + \chi_{\log L}^2 \right),
\label{likelihood_function}
\end{equation}
which is probed by sampling the posterior probability distribution during the MCMC routine. Each $\chi^2$ is defined as
\begin{align}
\chi_{\log T_{\mathrm{eff}}}^2 &= \frac{\left( \log T_{\mathrm{eff, obs}} - \log T_{\mathrm{eff, model}} \right)^2}{\sigma_{\log T_{\mathrm{eff}}}^2} \, \mathrm{and} \\
\chi_{\log L}^2 &= \frac{\left( \log L_{\mathrm{obs}} - \log L_{\mathrm{model}} \right)^2}{\sigma_{\log L}^2}.
\label{chi_square_function}
\end{align}

Since a given position in the HRD can correspond to multiple combinations of $M_\mathrm{core}$, $M_\mathrm{envelope}$, and age, the resulting posterior parameters are often degenerate. To reduce this degeneracy and improve sampling efficiency, we impose priors informed by each object’s spectroscopic type. 
For the sdB stars, a prior proportional to the inverse of the evolutionary speed in the HRD was applied, favouring slower evolutionary phases. Conversely, for sdO stars, the prior is proportional to the evolutionary speed itself, thereby favouring faster evolutionary phases consistent with their likely post-EHB evolutionary status. A flat probability distribution was assumed for the sdOB stars as their specific evolutionary status is unclear. 
No other priors were applied. 

The derived theoretical mass distributions are shown as KDEs of the entire posterior sample distribution in Fig. \ref{kde_mass_luminosity_mcmc}, where the central and left-most panels may be directly compared to the parallax-based mass distributions in Fig.\ \ref{kde_mass_luminosity} above.  
As shown in Table \ref{median_masses}, both the HRD and parallax-based masses are in excellent agreement. 
Moreover, the below EHB objects shown as dashed-red lines in the central panels also agree well with each other, both exhibiting a pronounced peak that is lower compared with the full sample. 
Because luminosity can be measured more precisely than \logg\ from spectroscopy, the theoretical masses exhibit narrower distributions. However, it should be noted that uncertainties in the evolutionary tracks themselves are not accounted for here. 

\begin{figure*}
    \centering
    \begin{subfigure}[b]{0.495\linewidth}
        \centering
        \includegraphics[width=\linewidth]{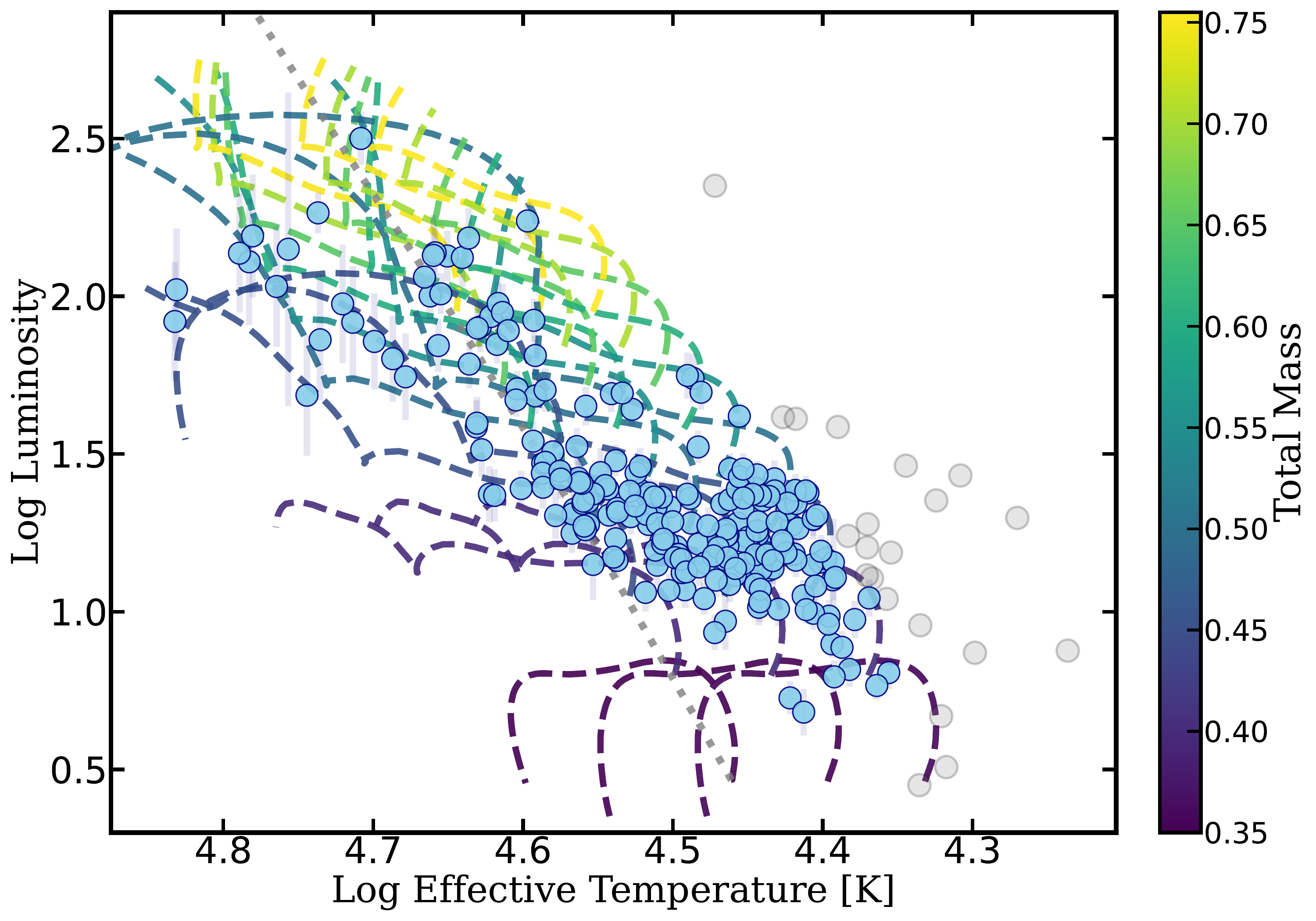}
        \label{track_interpolation_han}
    \end{subfigure}
    \hfill
    \begin{subfigure}[b]{0.48\linewidth}
        \centering
        \includegraphics[width=\linewidth]{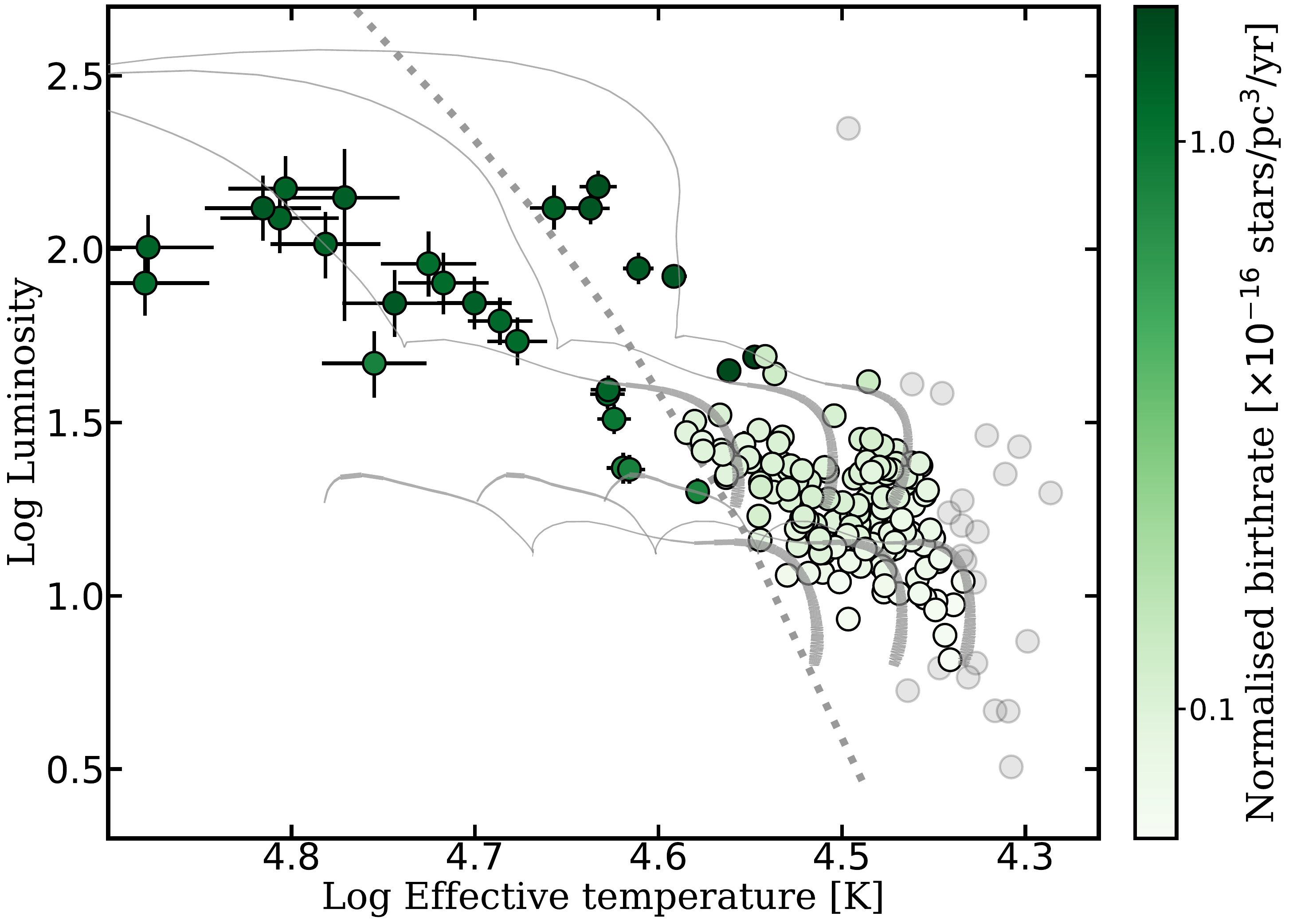}
        \label{track_interpolation_driebe}
    \end{subfigure}
    \caption{Left: The 253 single-lined hot subdwarf stars shown in the HRD, overlaid with the full set of evolutionary tracks from \citet{Han_2002}, colour-coded by increasing core mass as indicated by the colour bar. The interpolation and MCMC procedure used to estimate theoretical stellar masses and ages is described in Sect.~\ref{interpolation}. Faded points lie outside the interpolated region and are therefore excluded from mass determination. Right: Same as in the left panel but displaying only the 0.40 and 0.50 \msun\ tracks from \citet{Han_2002}, and the 183 stars that fit to them. The colour scale shows our expected birthrate calculations as outlined in Sect. \ref{birthrates}.}
    \label{fig:han_and_canonical}
\end{figure*}

The HRD-derived mass distribution seen for the sdOs is quite broad (0.41 to 0.65\,\msun; 68\% confidence). This is caused by their position in the HRD at high luminosities and temperatures, where there is a large overlap between the high-mass and the evolved lower-mass tracks. 
This degeneracy was partially alleviated by imposing a prior that favours more evolved positions along the tracks, although in some cases younger and more massive solutions remain preferred. The uncertainties on our parallax-based masses for sdOs (approximately $\pm 0.15\,M_\odot$) again prevent a definitive resolution of this degeneracy.

\subsection{Mass and luminosity distribution for He-sdO stars}

In the case of the eHe-sdO and iHe-sdOB stars, both observational and theoretical masses lie within the range of what is expected for the scenario involving the merging of two He-WDs \citep[0.4 to 0.7\,\msun; e.g.][]{Webbink_1984ApJ...277..355W, Saio_2000MNRAS.313..671S, Han_2003}. In particular, the HRD distribution of eHe-sdO masses ($0.52^ {+0.08}_{-0.07}$ \msun; Table \ref{median_masses}) is very similar to the predictions of \citet{Han_2003} for the merger channel (see their figure 12; $\approx$\,$0.54\pm 0.06$\,\msun). The slightly wider mass range seen in our HRD masses might point to a wider range of masses of the double white dwarf binaries than assumed in \citet{Han_2003}, or hint at contribution from other merger scenarios. 
Note that the models of \citet{Han_2003} did not consider helium-enriched envelopes and did not model the merger itself, limiting the accuracy of interpolated masses for He-rich stars.

The luminosity distribution of our eHe-sdO sample has a median of $\log (L/L_\odot) = 1.88^ {+0.24}_{-0.23}$, which is broadly consistent with the predicted median for the hybrid CO+He WD merger channel \citep{Justham_2010}. Our results somewhat favour the model variant in which approximately $0.1\,M_\odot$ of the He-WD is not accreted. This finely tuned channel, involving mergers between post-sdB and He WDs, also predicts a high-luminosity tail up to $\log (L/L_\odot) \approx 3.0$, and almost no objects with $\log (L/L_\odot) < 1.6$. In contrast, only one star in our sample exceeds $\log (L/L_\odot) > 2.5$, while we observe a clear extension of sources down to $\log (L/L_\odot) \approx 1.0$. Although the absence of a pronounced high-luminosity wing could be attributed to small-number statistics (our sample contains only 15 eHe-sdOs), the low-luminosity tail is not reproduced by the models of \citet{Justham_2010}. Given that our sample includes a non-negligible fraction ($\sim10 \%$) of low-mass, low-luminosity sdBs, this may offer a constraint on the input merger population for the models.

Our population of iHe-sdOBs divides into two distinct groups. The lower-luminosity objects coincide with the He-poor sdOBs, whereas the higher-luminosity objects are more closely associated with the eHe-sdOs in both the HR and Kiel diagrams. A similar division was also reported by \citet{Dorsch_2024PhDT........36D} in their SED analysis of the spectroscopically identified hot-subdwarf sample of \citet{Culpan_2021}. 

The evolutionary tracks of \citet{Han_2002} were computed only for core masses up to 0.75 \msun, and therefore may not cover several of the more luminous iHe-sdOB and eHe-sdO stars (see the proximity of these stars to the He-MS in the HRD in Fig.\ \ref{hrd}). Since all our helium-rich objects lie close to the He-MS of \citet{Paczynski1971} in the HRD, we estimated the masses of all He-rich stars in our sample by projecting their positions onto this sequence. The resulting mass distributions, shown in the right panel of Fig.\ \ref{kde_mass_luminosity_mcmc} along with the interpolated masses from \citet{Han_2002}, estimate the iHe-sdOB stars to peak at $0.59^ {+0.08}_{-0.07}$\,\msun, while the eHe-sdO stars show a broader distribution between 0.55 and 0.85\,\msun\ (68\,\% range). These values are systematically higher than those inferred from the parallax in the SED method and the \citet{Han_2002} tracks, and would therefore require the progenitor population of close double white dwarfs to span a similarly broad mass range. For TYC\,5720-292-1 (annotated in Fig.~\ref{mass_temperature}), the parallax-based and He-MS masses are in good agreement; however, for BD$+$39\,3226, the He-MS mass is about 0.3 \msun\ lower. 

\subsection{Intermediate-mass progenitors}
\label{intermediate_mass_progenitors}

Hot subdwarfs are commonly discussed in the canonical framework of low-mass progenitors (about $0.7-1.8\, M_{\odot}$), in which helium ignition occurs via a core flash at a mass of roughly half a solar mass. This picture is supported by a corresponding peak in the observed mass distribution \citep[][also Sect.\ \ref{mass_distribution}]{Schaffenroth_2022_1, Lei_2023ApJ...953..122L_mass}. 
Intermediate-mass stars ($1.8-8.0\, M_{\odot}$) ignite helium smoothly (non-degenerately or partially degenerately), without a helium flash, forming cores as low as $\sim0.3$ \msun, up to or more than $1.0\,M_{\odot}$ \citep{Cox_1964ApJ...140..485C,Hansen_1971ApJ...163..653H,Han_2002,Arancibia_Rojas_2024MNRAS.52711184A}\footnote{See their fig.\ 4  for an illustration of the relation between He-core masses and the masses of the progenitor stars on the main sequence.}. 

We identify a distinct population of low-mass, low-luminosity sdBs, which account for about 10\,\% of EHB stars in our sample. This population is supported by both observational evidence (Sect.\ \ref{low_lum_sdbs}, \ref{low_mass_sdbs}) and theoretical models (Sect. \ref{interpolation}) regarding their masses and HRD positions, and we interpret them as descendants of intermediate-mass progenitors.

The prevalence of such stars in our sample likely reflects the fact that it is dominated by thin-disk objects (Sect.~\ref{kinematics}); comparatively younger intermediate-mass progenitors are not expected in the older thick disk or the halo because of the lack of star formation there and we do not find the corresponding low-luminosity sdBs in those populations either (see Fig. \ref{hrd_thin_thick_halo}), though it must be stressed that our sample contains only 22 thick disk stars and 4 halo stars. 
In particular, producing $\sim$0.40\,$M_{\odot}$ sdBs requires progenitors in a fairly tight initial mass range of about $1.8$ to $3.0\,M_{\odot}$. 
The formation of sdBs with masses up to and above $1.0\,M_{\odot}$ is possible; however, their core helium-burning lifetimes are short  \citep{Yungelson_2008AstL...34..620Y, Arancibia_Rojas_2024MNRAS.52711184A, Nicolas_2_2025PASA...42...12R}. 
Our parallax-based mass estimates yielded several sdB and sdOB stars with masses between 0.6 and $0.7\,M_{\odot}$ (Fig. \ref{mass_temperature}) which may descend from progenitors exceeding $4\,M_{\odot}$. However, given the associated uncertainties and their positions on the EHB in the HRD (Fig.\ \ref{hrd}), the nature of their progenitors remains ambiguous. 
We therefore focus on the low-mass sdB population, which can clearly be identified as descendants of intermediate-mass stars. 

Several binary interaction channels are able to produce hot subdwarfs from intermediate-mass progenitors \citep{Han_2003, Nicolas_2_2025PASA...42...12R}. 
These include the first stable RLOF in the Hertzsprung gap or on the subgiant branch, CE stripping near the tip of the RGB (first CE channel), and CE stripping by a WD companion (second CE channel).
In each case, low-mass hot subdwarfs are produced by progenitors more massive than around 1.9 \msun, a transitional point between degenerate and non-degenerate helium ignition \citep[see][for more details]{Han_2003}.

Recent theoretical work by \cite{Nicolas_1_2025MNRAS.539.3273R} on the present-day population of hot subdwarfs produced a peak in the mass distribution at around 0.35\,\msun, which is dominated by first RLOF from  intermediate-mass stars (see their figures\ 2 and 7), also found by \citet{Han_2002}. 
The large majority of these low-mass hot subdwarfs are predicted to have MS companions with masses between 1 and 4\,\msun; such systems were not considered in the present analysis given their composite spectra and SEDs and will be investigated in a future paper. 

In the best-fitting model of \citet{Han_2003}, the mass distribution for the first CE-ejection channel (producing sdB + low-mass MS systems) shows a dominant peak at 0.46\,\msun\ (degenerate ignition) and a secondary peak at 0.40\,\msun\ (semi-degenerate ignition), the latter of which resembles the masses of our below-EHB stars.
Two analogous mass peaks are also predicted to be produced by the second CE channel, which leaves sdB + WD systems. 
For the second CE channel, a third much smaller peak is predicted at 0.33\,\msun, which is linked to second CE ejection in the Hertzsprung gap with non-degenerate ignition. 
In addition, \cite{Nicolas_1_2025MNRAS.539.3273R} found that among the 0.35\,\msun\ EHB stars formed through a second CE, the companions are predominantly CO-core WDs, whereas the canonical 0.47\,\msun\ EHB stars formed through a second CE typically have He-core WD companions. 
Distinguishing between first and second CE requires knowledge of the nature of the companions and the orbital periods after mass transfer. Therefore, the binary properties of all stars in the 500\,pc sample will be investigated in a forthcoming paper (Dawson et al.\ in prep.). 

\subsection{Diffusion in low-mass EHB stars}
\label{diffusion}

In Sect.\ \ref{stellar_params} we noted that roughly half of the underluminous sdBs have very low helium abundances (\logy\ $\leq -3.8$) and higher surface gravities, indicative of strong gravitational settling. Although the settling timescale is not well constrained, \citet{Byrne_2018MNRAS.475.4728B} show that helium depletion can occur rapidly, even before the EHB, and drop below observed levels. Including artificial envelope mixing prolongs helium evolution throughout the EHB \citep[beyond 30 Myr;][]{Michaud_2011A&A...529A..60M}. Diffusion can be moderated by weak stellar winds \citep{Fontaine_Chayer_1997fbs..conf..169F,Unglaub_2008A&A...486..923U, Hu_2011MNRAS.418..195H} or shallow surface convection zones \citep{Unglaub_2008A&A...486..923U}. 
Stars with longer evolutionary timescales may undergo more extended diffusion, which could explain the extremely low helium abundances observed in some underluminous sdBs. The lifetimes of hot subdwarfs do not scale linearly with mass; thus, if the stars located below the EHB are indeed of low mass ($\sim0.33$–$0.45$\msun), their EHB lifetimes could range between $\sim200$ and $600$ Myr according to the models of \citet{Han_2003}. While all stars with \logy\ $< -3.8$ lie below the EHB, not all below-EHB stars are helium-poor (Fig.~\ref{helium}), suggesting that variations in initial helium abundance, age, or mass may also contribute to the observed spread.

Evolutionary channels involving intermediate-mass progenitors are expected to preferentially produce hot subdwarfs with helium-rich envelopes at formation  \citep{Justham_2010,Brown_2001ApJ...562..368B,Xiong_2017A&A...599A..54X,Nicolas_2025}, but this does not translate into observable atmospheric abundances due to the diffusive nature of sdB atmospheres.

\subsection{Canonical EHB and post-EHB birthrates}
\label{birthrates}

Generally, strong agreement is seen between our parallax-based masses and the theoretical masses derived from HRD tracks \citep{Han_2002}.
However, because many of the tracks in the HRD overlap, the theoretical masses of several objects become degenerate, resulting in broad or even multi-modal distributions. 
Our goal in this section is to compare the birthrate of EHB and post-EHB stars; this is only possible when the stellar mass is well constrained. 
For this experiment, we therefore limited our selection to stars that lie between the 0.40 and 0.50 \msun\ tracks (see the right panel of Fig.\ \ref{fig:han_and_canonical}) and interpolate between them using the same method as described in Sect. \ref{interpolation}.
For this analysis, we exclude the helium-enriched stars (eHe-sdO and iHe-sdOB), which are likely merger products. The O(H) and O(He) stars are also omitted; however, since their effective temperatures are significantly higher than those of the 0.40 and 0.50 \msun\ evolutionary tracks, their exclusion has no effect on our results.
The final number of objects which fit to these tracks total 183 out of 211 H-rich stars and are shown as scattered circles. The objects outside of the interpolated region are shown as grey circles. 

To derive the birthrate, we require a space density, $\rho_{0}$, and estimated lifetime, $\tau$, of each object. 
In Paper I we modelled the vertical distribution of the full 500~pc sample and derived the scale height and mid-plane density of the population using a single-component $\mathrm{sech}^2$ density profile (see Paper I for details). 
Here we leverage our kinematic results and employ a three-component $\mathrm{sech}^2$ density profile, representing the thin disk, thick disk, and halo populations. Due to low number statistics, we do not fit the scale heights of each component and instead adopt the following: $z_{h,\mathrm{thin}} = 300$~pc, $z_{h,\mathrm{thick}} = 900$~pc, and $z_{h,\mathrm{halo}} = 2000$~pc, which is in line with previous determinations using large samples of stars \citep[see][]{Juric_2008ApJ...673..864J}. The relative contribution of each component is set by the fractional number of stars in our sample identified as belonging to each population (see Table \ref{population_membership}).

\begin{align}
\rho(z) = \frac{1}{N} \bigg[ &
    \rho_{0,\mathrm{thin}} \, \mathrm{sech}^2\left(\frac{z}{2 z_{h,\mathrm{thin}}}\right) \notag \\
    &+ \rho_{0,\mathrm{thick}} \, \mathrm{sech}^2\left(\frac{z}{2 z_{h,\mathrm{thick}}}\right) \notag \\
    &+ \rho_{0,\mathrm{halo}} \, \mathrm{sech}^2\left(\frac{z}{2 z_{h,\mathrm{halo}}}\right)
\bigg]
\label{rho_profile}
\end{align}
\noindent
Equation~\ref{rho_profile} gives the normalised stellar number density as a function of vertical height $z$ above the Galactic plane. We then scale the profile vertically by fixing the function to the mid-plane density derived in Paper I: $\rho_{0} = 5.17 \pm 0.33 \times 10^{-7}$ stars pc$^{-3}$, thereby ensuring that the profile reflects an absolute space density. Equation \ref{rho_profile} therefore represents the expected stellar density (in units of $\text{stars} / \text{pc}^3$) at position $z$, and can be used to assign an expected space density for each star based on its position $z$.

For each star $i$, we determine the age of the best-fit track at the point of core helium exhaustion, $\tau_{\mathrm{EHB},i}$, which defines the transition from the EHB to the post-EHB phase. From this, we can separate the lifetimes of the two stages: the EHB lifetime is $\tau_{\mathrm{EHB},i}$, and the post-EHB lifetime is the remaining time, $\tau_{\mathrm{pEHB},i} = \tau_i - \tau_{\mathrm{EHB},i}$, with $\tau_i$ denoting the total lifetime.

Having estimated both the local space density and evolutionary lifetime for each object, we divide the expected space density $\rho_i$ by the corresponding phase lifetime ($\tau_{\mathrm{EHB},i}$ or $\tau_{\mathrm{pEHB},i}$) obtained from the tracks. This yields an expected birthrate for each of the 183 objects in this experiment, based on their Galactic $z$-position (through the model-predicted local density) and evolutionary phase. Although not a direct physical birthrate, this quantity traces the relative contribution of stars with different ages and vertical positions to the overall population.

The total estimated birthrates for the EHB and post-EHB stars are derived as
\begin{itemize}
  \item[] $\eta_\mathrm{EHB} = \sum_i \rho_{i} / \tau_{\mathrm{EHB},i} = 1.41\substack{+0.01\\ -0.02} \times 10^{-15}$ stars/pc$^{3}$/yr$^{-1}$ and \vspace{2pt}
  \item[] $\eta_\mathrm{pEHB} = \sum_i \rho_{i} / \tau_{\mathrm{pEHB},i} = 3.30\substack{+0.17\\ -0.10} \times 10^{-15}$ stars/pc$^{3}$/yr$^{-1}$.
\end{itemize}
This may be interpreted as the birthrate of the post-EHB objects being 2-3 times larger than the EHB objects. 
There could be several reasons for this discrepancy. 
One possibility is that the post-EHB region includes H-rich stars that are not actually in the post-EHB phase, such as more massive sdBs originating from intermediate-mass progenitors, whose core-helium burning occurs at luminosities above the 0.5\,\msun\ EHB, though their lifetimes would be short. 
Three sdB stars were removed due to high RUWE values (Sect. \ref{low_mass_sdbs}), and several stars lie higher up on the EHB beyond the tracks (grey circles), but alone they are not sufficient to explain the 2-3 fold discrepancy because if on the EHB they would count less towards the birthrate than post-EHB objects would due to their longer phase lifetimes. 
Another possibility is that the EHB lifetimes predicted by the evolutionary tracks of \citet{Han_2002} are systematically overestimated, which would directly affect the relative birthrate ratio of EHB to post-EHB stars derived above. Compared with other models, the \citet{Han_2002} tracks begin with a higher central helium mass fraction, averaging about $0.973$ between the 0.40 and 0.50\,\msun\ sequences, which corresponds to an EHB lifetime of about $180$\,Myr. For comparison, the 0.475\,\msun\ solar metallicity models from \citet{Dorman_1993ApJ...419..596D_models} start with a lower central helium fraction of 0.95, yielding a shorter EHB lifetime of $\sim136$\,Myr. However, this modest 1-2\% reduction in the available helium fuel would account for only a few Myr difference in the evolutionary timescales of the \citet{Han_2002} models. Thus, the substantially longer lifetimes are more plausibly attributed to intrinsic differences in the model physics rather than variations in the initial helium content.
Another possible source of inconsistency in the estimated occurrence rates of EHB and post-EHB stars is that our assumed mass range of 0.40 to 0.50\,\msun\ may be too narrow, which is likely (see Sect.\ \ref{mass_distribution}). 

In addition, some low-mass post-EHB stars at $\log L / L_\odot > 1.05$ may currently be misclassified as EHB stars, which would also exacerbate the discrepancy. 
More massive EHB stars could also contribute to the post-EHB region, but this is unlikely, as our parallax-based masses exclude values above 0.7 \msun\ at 68\% confidence in the range between the 0.40 and 0.50 \msun\ tracks (Fig.~\ref{mass_temperature}).

Finally, helium-core white dwarfs evolving towards the cooling sequence are predicted to occupy the same region of parameter space as analysed here \citep{Driebe_1999ASPC..169..394D, Istrate_2016A&A...595A..35I, althaus2025postcommonenvelopeevolutionheliumcore}, which may also contribute to the observed birthrate discrepancy. Indeed, we identify CPD-20\,1123 as a likely candidate, along with the previously known EVR-CB-001. A more detailed investigation of helium-core white dwarfs within this region of parameter space will be presented in a forthcoming paper.

\section{Conclusions and outlook}

This work extends the spectroscopic study of the 500 pc sample of single-lined hot subdwarf stars introduced in Paper I, providing for the first time a homogeneous analysis of a volume-complete sample. We analysed 3226 spectra of 253 objects by fitting model atmospheres to derive \teff, \logg, \logy, and $v_\mathrm{rad}$ for each star. We further used a subset of our spectra to assess systematic uncertainties in the derived atmospheric parameters. These uncertainties, modelled with third-order polynomial fits, can be added in quadrature to future statistical errors.
Stellar masses, radii, and luminosities were determined by combining spectroscopic parameters with SED fits and \textit{Gaia} parallaxes. The resulting mass distributions and HR diagrams were quantitatively compared with theoretical evolutionary tracks. Finally, using the measured $v_\mathrm{rad}$ and inferred systemic velocities, we performed a kinematical analysis to determine the Galactic population membership of each object.
Our analysis allowed us to draw several conclusions:

   \begin{enumerate}
      \item A significant portion of the observed sample ($10\pm 2$\,\%; binomial uncertainty) likely originates from intermediate-mass progenitors ($1.8\,-\,8\,M_\odot$), as evidenced by the prominent population of hot subdwarfs situated below the EHB in both the Hertzsprung-Russell and Kiel diagrams. These systems also have, on average, lower masses compared to the general population which is corroborated by theoretical models. This points to a potentially greater contribution from the non-degenerate, no-helium flash formation channel within the Galactic disk than has traditionally been assumed, and encourages future studies in the intermediate-mass regime. 
      A consequence of the significant contribution of low-mass EHB stars implies that assuming canonical masses for hot subdwarfs in single-lined binaries is not reliably justified.
      \vspace{0.5em}
      \item We identify two groups of sdBs with extremely helium poor atmospheres (\logy\ $\leq$ $-3.8$) in the temperature range of $20\,000$ to $40\,000$\,K. The cool sdBs ($T_\mathrm{eff} < 28\,000$\,K) are almost exclusively located below the EHB, whereas the hotter group ($T_\mathrm{eff} > 34\,000$\,K) is positioned above it, establishing a possible evolutionary link between the cool helium-poor sdBs and the sdO stars. On average, these stars have lower masses (0.41 \msun) and therefore longer lifetimes. This may allow diffusive effects in the atmospheres in these stars to operate for extended periods, or mixing processes to weaken, leading to lower helium abundances at the stellar surface. 
        \vspace{0.5em}
      \item Our kinematic analysis, based on astrometry from \textit{Gaia} and the radial velocities from 3226 spectra of 253 hot subdwarfs, shows that $86\pm2 \%$  belong to the Galactic thin disk, while $13 \pm1 \%$ and $1 \pm 1 \%$ belong to the thick disk and halo populations, respectively. This analysis further revealed that the hot subdwarf stars situated below the EHB are exclusive to the thin disk, which is again consistent with their proposed origin from young intermediate-mass stars.
      \vspace{0.5em}
      \item We compared our masses, derived from SED fits using \textit{Gaia} parallaxes, to theoretical masses calculated by interpolating within the evolutionary tracks from \citet{Han_2002} using an MCMC procedure, which returned a strong agreement between them. The HRD-based mass distributions are also narrower than those derived from the \textit{Gaia} parallax in the SED fits likely owing to the precise luminosities. 
      However, both methods carry significant uncertainties: HRD fitting is affected by degeneracies from overlapping tracks and the choice of models, while parallax-based masses are sensitive to uncertainties in spectroscopic \logg\ measurements.
      \vspace{0.5em}
      \item By assuming a restricted mass range for our stars (0.40 - 0.50 \msun) and interpolating the corresponding tracks only, we find that the post-EHB birthrate would be 2-3 times higher than the EHB birthrate when adopting a Galactic profile space density for our population based on the results from paper I. A possible reason for the discrepancy is an overestimation of the EHB lifetimes in the tracks of \citet{Han_2002}. However, other sources of bias and contamination such as the inclusion of H-rich stars in the post-EHB region, which are actually in the EHB phase, cannot be ruled out at this stage. 
   \end{enumerate}

We provide atmospheric, stellar, and kinematic properties of hot subdwarfs using the first volume-complete sample designed to minimise observational biases. This offers valuable constraints for the next generation of binary population synthesis models. Building on this foundation, forthcoming studies (Dawson et al., in prep.) will present a detailed characterisation of the close binary population within 500~pc, while the 48 wide systems with identified A/F/G-type companions from Paper~I will be analysed separately. Just 4 stars within 500~pc are classified as halo objects (Sect. \ref{kinematics}), three of them being helium-enriched and the forth a BHB. Therefore, looking ahead, the next major step is to extend volume-complete samples to different galactic environments to incorporate more thick disk and halo stars where we expect to find e.g. a larger fraction of post-merger objects. A new survey targeting the Galactic north pole is nearing completion and will enable a direct comparison of population properties across galactic components in an unbiased way for the first time, providing fresh insight into the role of environment in hot subdwarf formation and evolution.

\section{Data availability}
\label{sect:data_availability}
The full versions of Tables \ref{main_table_atm}, \ref{main_table_sed}, and \ref{main_table_kin} are only available in electronic form at the CDS via anonymous ftp to \url{cdsarc.u-strasbg.fr (130.79.128.5)} or via \url{http://cdsweb.u-strasbg.fr/cgi-bin/qcat?J/A+A/}. 

\begin{acknowledgements}

We thank Philipp Podsiadlowski for useful discussions. 
H. D. is and N. R. was supported by the Deutsche Forschungsgemeinschaft (DFG) through grant GE2506/17-1. 

M.\ D.\ was supported by the Deutsches Zentrum für Luft- und Raumfahrt (DLR) through grant 50-OR-2304. 

N.R. is supported by the Deutsche Forschungsgemeinschaft (DFG) through grant RE3915/2-1.

V. S. and M. P. received funding by the Deutsche Forschungsgemeinschaft (DFG) through grants GE2506/9-1 and GE2506/12-1.
D.S. acknowledges funding by DFG grant HE1356/70-1.

I. P. acknowledges funding from a Warwick Astrophysics prize post-doctoral fellowship, made possible thanks to a generous philanthropic donation, and from a Royal Society University Research Fellowship (URF\textbackslash R1\textbackslash 231496).

J.V. acknowledges support from the Grant Agency of the Czech Republic (GA\v{C}R 22-34467S). The Astronomical Institute Ond\v{r}ejov is supported by the project RVO:67985815.

M. U. gratefully acknowledges funding from the Research Foundation Flanders (FWO) by means of a junior postdoctoral fellowship (grant agreement No. 1247624N).

K.D. acknowledges funding from the Methusalem grant METH/24/012 at KU Leuven. This research has used observations obtained at the Mercator Observatory which receives funding from the Research Foundation – Flanders (FWO) (grant agreement I000325N and I000521N).

A.B. was supported by the Deutsche Forschungsgemeinschaft (DFG) through grant GE2506/18-1. 

T. S. acknowledges funding from grant SONATA BIS no 2018/30/E/ST9/00398 from the Polish National Science Centre (PI T. Kami\'{n}ski).

R.R. acknowledges support from Grant RYC2021-030837-I funded by MCIN/AEI/ 10.13039/501100011033 and by “European Union NextGeneration EU/PRTR”. This work was partially supported by Spanish MINECO grant PID2023-148661NB-I00.

T.K. acknowledges support from the National Science Foundation through grant AST \#2107982, from NASA through grant 80NSSC22K0338 and from STScI through grant HST-GO-16659.002-A. Co-funded by the European Union (ERC, CompactBINARIES, 101078773). Views and opinions expressed are however those of the author(s) only and do not necessarily reflect those of the European Union or the European Research Council. Neither the European Union nor the granting authority can be held responsible for them.

M.L. acknowledges funding from the DFG (grant LA 4383/4-1).

This project has received funding from the European Research Council under the European Union’s Horizon 2020 research and innovation programme (Grant agreement numbers 101002408).

Based on observations collected with the Goodman spectrograph at the Southern Astrophysical Research Facility (SOAR) at Cerro Pachon, Chile, under the programme allocated by the Chilean Telescope Allocation Committee (CNTAC), no: 2023B, 2024A and 2025A.

This work has made use of the BeSS database, operated at LESIA, Observatoire de Meudon, France: http://basebe.obspm.fr. 
Some data in this worked came from Guaranteed Observation Time (GTO) based on observations collected at the Centro Astronomico Hispano en Andalucia (CAHA) at Calar Alto, operated jointly by Junta de Andalucia and Cosejo Superior De Investigaciones Cientificias (IAA-CSIC). The research has made use of TOPCAT, an interactive graphical viewer and editor for tabular data \citep{TOPCAT_2005ASPC..347...29T}. This research made use of the SIMBAD database, operated at CDS, Strasbourg, France; the VizieR catalogue access tool, CDS, Strasbourg, France. 
This work has made use of data from the European Space Agency (ESA) mission \textit{Gaia} (https://www.cosmos.esa.int/gaia), processed by the \textit{Gaia} Data Processing and Analysis Consortium (DPAC, https://www.cosmos.esa.int/web/gaia/dpac/consortium). Funding for the DPAC has been provided by national institutions, in particular the institutions participating in the \textit{Gaia} Multilateral Agreement.

\end{acknowledgements}

\bibliography{Bibliography}
\bibliographystyle{aa}

\begin{appendix}

\section{Additional tables and figures}
\label{appendix_a}

\begin{table}[]

\caption{Included hot subdwarfs with a hint of IR excess.}
\label{IR_excess}
\renewcommand{\arraystretch}{1.2}
\centering
        \begin{tabular}{ccc}
                \toprule\toprule
                Name & Class & Reference    \\
                \midrule 
                 TYC 7489-686-1 & sdB+dM/BD & [6, 11]  \\
                 Feige 34 & sdO+dM & [7] \\
                 V*EQPsc/PB 5450 & sdBV+dM & [5]  \\
                 PG 1619+522 & sdB+dM & [3, 4]  \\
                 GD 1068 & sdB+dM & [10]  \\
                 CD$-$23 15853 & sdO+dM & This work.  \\
                 Feige 36 & sdB+dM & [1, 2]  \\
                 HD 149382$^\ast$ & sdOB+dM & [8, 12, 13, 14]  \\
                 TYC 5977-517-1 & sdB+dM/BD & [9]  \\
                 UCAC4 219-125136$^\dag$ & sdB+? & This work.\\
                \bottomrule
        \end{tabular}
\tablefoot{
$^\ast$ The hint of IR excess in the SED more likely comes from a nearby contamination star.\\
$^\dag$ A strong IR excess in this system likely comes from background contamination. \\ 
\textbf{References.} 1) \citet{Saffer_1998ApJ...502..394S}, 2) \citet{Moran_1999MNRAS.304..535M}, 3) \citet{Maxted_2001}, 4) \citet{Morales_2003MNRAS.338..752M}, 5) \citet{Jeffery_2014MNRAS.442L..61J}, 6) \citet{Kawa_2015}, 7) \citet{Latour_2018AA...609A..89L}, 8) \citet{Schneider_2018}, 9) \cite{Krzesinski_2022yCat..36630045K}, 10) \cite{Schaffenroth_2022_1} 11) \citet{Schaffenroth_2023_2}, 12) \cite{Geier_2009ApJ...702L..96G}, 13) \citet{Ostensen_2005ASPC..334..435O}, 14) \citet{Jacobs_2011AIPC.1331..304J}
}
\end{table}

\begin{table}[h]
\centering
\footnotesize
\caption{Coefficients of the third-order polynomial fits to the systematic uncertainty for each parameter as in Eq.~\ref{uncertainty_polynomial}. The coefficients for \teff\ are given in kilokelvin.}
\label{tab:uncertainty_coeffs}
\setlength{\tabcolsep}{4pt} 
\renewcommand{\arraystretch}{1.0} 
\begin{tabular}{lrrrr}
\toprule
\toprule
Parameter & $a_3$ & $a_2$ & $a_1$ & $a_0$ \\
\midrule
      &  \multicolumn{4}{c}{He-poor} \\
\midrule
$T_{\mathrm{eff}}$  & 1.09$\times$10$^{-6}$  & -5.37$\times$10$^{-5}$ & 4.13$\times$10$^{-4}$ & 1.68$\times$10$^{-2}$ \\
$\log g$            & -2.75$\times$10$^{-6}$ & 3.11$\times$10$^{-4}$  & -1.07$\times$10$^{-2}$ & 1.79$\times$10$^{-1}$ \\
\logy\            & -1.64$\times$10$^{-5}$ & 2.08$\times$10$^{-3}$  & -8.17$\times$10$^{-2}$ & 1.07$\times$10$^{0}$ \\
\midrule
      &  \multicolumn{4}{c}{He-rich} \\
\midrule
$T_{\mathrm{eff}}$  & 1.47$\times$10$^{-5}$  & -1.73$\times$10$^{-3}$ & 6.84$\times$10$^{-2}$  & -8.91$\times$10$^{-1}$ \\
$\log g$            & 2.68$\times$10$^{-5}$  & -3.30$\times$10$^{-3}$ & 1.37$\times$10$^{-1}$  & -1.82$\times$10$^{0}$ \\
\logy\            & -6.92$\times$10$^{-5}$ & 8.49$\times$10$^{-3}$  & -3.40$\times$10$^{-1}$ & 4.50$\times$10$^{0}$ \\
\bottomrule
\end{tabular}
\end{table}

\begin{figure}
    \centering
    \begin{subfigure}[b]{\linewidth}
        \centering
        \includegraphics[width=\linewidth]{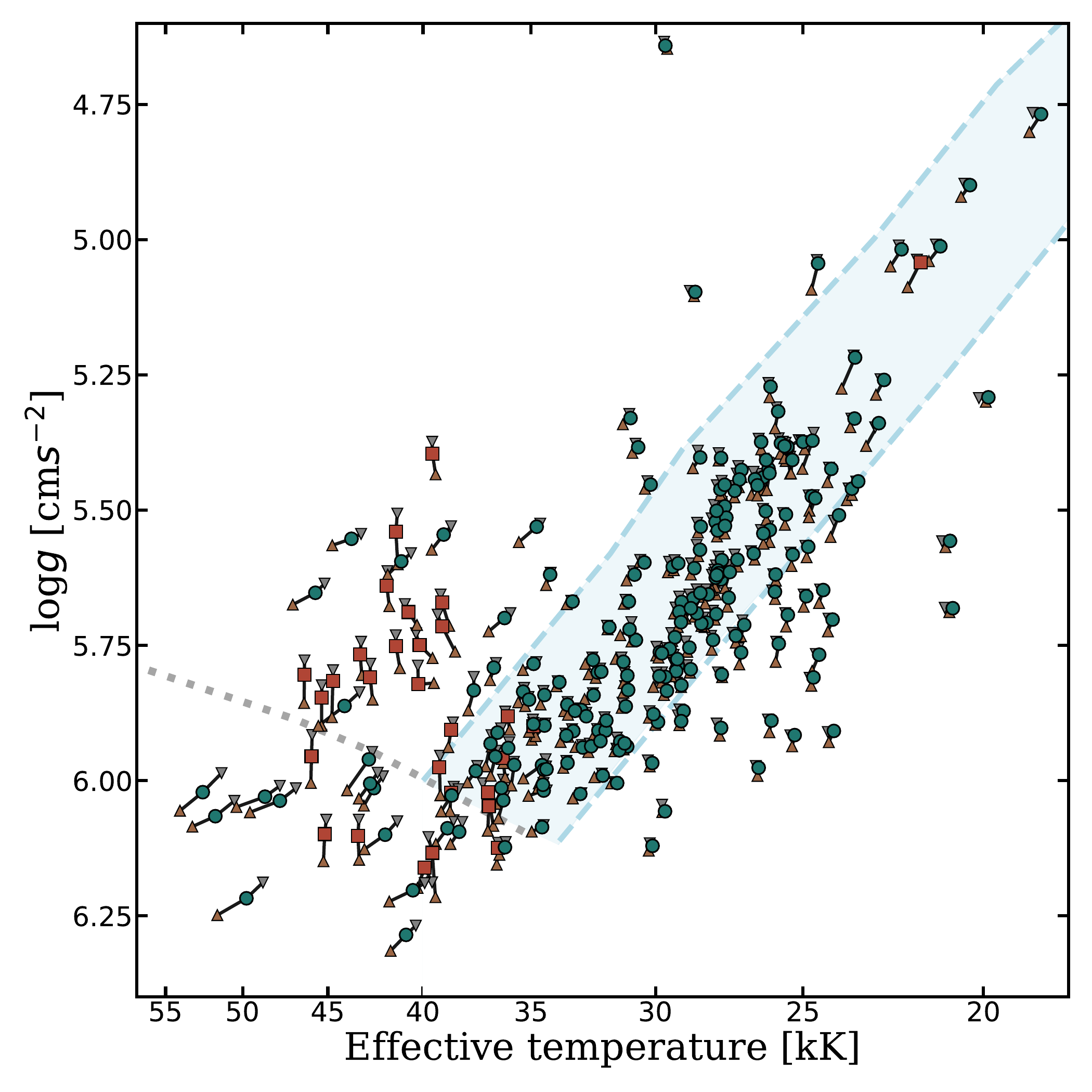}
        \caption{}
        \label{kiel_z}
    \end{subfigure}
    \begin{subfigure}[b]{\linewidth}
        \centering
        \includegraphics[width=\linewidth]{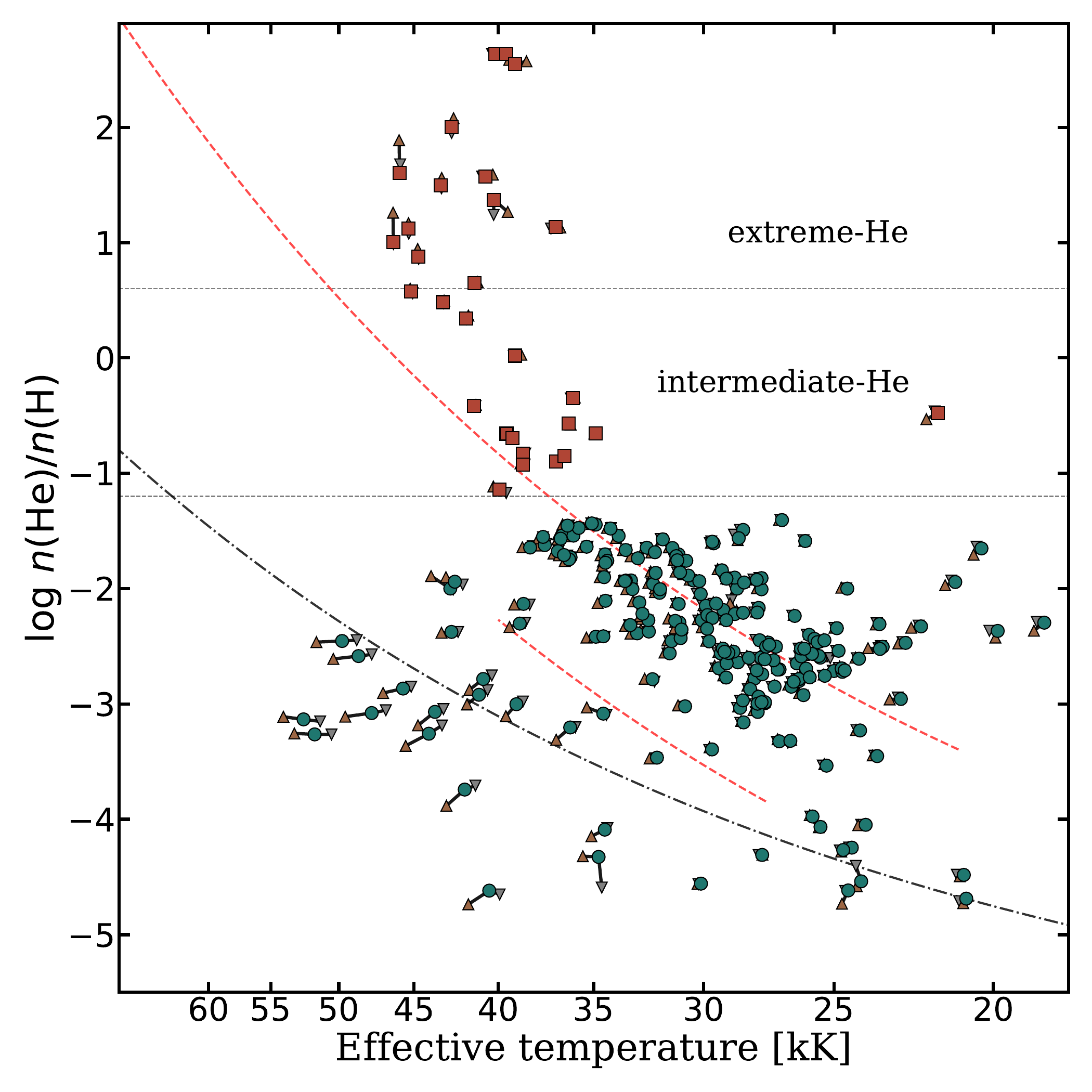}
        \caption{}
        \label{helium_z}
    \end{subfigure}
    \caption{\logg\ versus effective temperature (top) and helium abundance versus effective temperature (bottom) showing the impact of metallicity in these parameter spaces. Circles and squares mark the helium-poor (\logy $<$ –1.2) and helium-rich (\logy $>$ –1.2) stars, respectively, for fits fixed to a metallicity of $\log z / z_\mathrm{sdB}$ $=0$. Grey and bronze triangles, oriented such that they point down and up, respectively, show the parameters derived for these same stars when fixing the metallicity to $\log z / z_\mathrm{sdB}$ $= -0.3$ and $+0.3$, respectively. Each triangle is connected by a black line to the circle or square to indicate the shift.}
    \label{fig:metallicity}
\end{figure}

\end{appendix}
\end{document}